\documentclass[iop]{emulateapj}

\newcommand{\HI}{\mathrm{H\,I}}

\newcommand{\GHI}{\Gamma_{\HI}}

\newcommand{\mfp}{$\lambda_{\mathrm{mfp}}$}

\newcommand{\lya}{Ly$\alpha$ }
\newcommand{\lyb}{Ly$\beta$ }
\newcommand{\lyg}{Ly$\gamma$ }
\newcommand{\lyam}{{\mathrm{Ly}\alpha}}
\newcommand{\lybm}{{\mathrm{Ly}\beta}}
\newcommand{\lygm}{{\mathrm{Ly}\gamma}}
\newcommand{\lyab}{Ly$\alpha$+$\beta$}

\newcommand{\teff}{\tau_\mathrm{eff}}

\begin{document}
\title{A New Method to Measure the Post-Reionization Ionizing Background from the Joint Distribution of Lyman-$\alpha$ and Lyman-$\beta$ Forest Transmission\footnotemark[*]}

\footnotetext[*]{Some of the data presented herein were obtained at the W.M. Keck Observatory, which is operated as a scientific partnership among the California Institute of Technology, the University of California and the National Aeronautics and Space Administration. The Observatory was made possible by the generous financial support of the W.M. Keck Foundation.}

\author{Frederick B. Davies$^{1}$, Joseph F. Hennawi$^{2,1}$, Anna-Christina Eilers$^{1,}$\footnotemark[$\dagger$], Zarija Luki\'{c}$^3$}
\affil{$^1$Max-Planck-Institut f{\"u}r Astronomie, K{\"o}nigstuhl 17, D-69117 Heidelberg, Germany\\
$^2$Department of Physics, University of California, Santa Barbara, CA 93106, USA\\
$^3$Lawrence Berkeley National Laboratory, CA 94720-8139, USA}
\email{davies@mpia.de}
\footnotetext[$\dagger$]{Fellow of the International Max Planck Research School for Astronomy and
Cosmic Physics at the University of Heidelberg (IMPRS-HD).}

\begin{abstract}
  The amplitude of the ionizing background that pervades the
  intergalactic medium (IGM) at the end of the epoch of reionization
  provides a valuable constraint on the emissivity of the sources
  which reionized the Universe. While measurements of the ionizing
  background at lower redshifts rely on a simulation-calibrated
  mapping between the photoionization rate and the mean transmission
  of the \lya forest, at $z\gtrsim6$ the IGM becomes increasingly
  opaque, and transmission arises solely in narrow spikes separated by
  saturated Gunn-Peterson troughs. In this regime, the traditional approach of
  measuring the average transmission over large $\sim 50$ Mpc$/h$ regions
  is less sensitive and sub-optimal. Additionally, the five times
  smaller oscillator strength of the Ly$\beta$ transition implies the \lyb forest is
  considerably more transparent at $z\ga6$, even in the
  presence of contamination by foreground $z\sim 5$ Ly$\alpha$ forest absorption. 
  In this work we present a novel statistical approach to analyze the joint
  distribution of transmission spikes in the co-spatial $z\sim 6$ Ly$\alpha$ and Ly$\beta$ forests. 
  Our method relies on Approximate Bayesian Computation (ABC), which circumvents
  the necessity of computing the intractable likelihood function describing the
  highly correlated Ly$\alpha$ and Ly$\beta$ transmission. We apply ABC to mock data generated from a large-volume hydrodynamical simulation combined with a state-of-the-art model of ionizing background fluctuations in the post-reionization IGM, and show that it is sensitive to higher IGM neutral hydrogen fractions than previous
  techniques. As a proof of concept, we apply this methodology to a real spectrum of a $z=6.54$ quasar and
  measure the ionizing background from $5.4\leq z \leq 6.4$ along this sightline with $\sim0.2$ dex statistical
  uncertainties.

\end{abstract}

\section{Introduction}

Following the recombination of the Universe and the release of the cosmic microwave background (CMB), the vast majority of baryonic matter in the intergalactic medium (IGM) consisted of neutral atoms, in stark contrast to the highly ionized IGM seen a few billion years later \citep{GP65} and at the present day \citep{Field59}. The first stars, galaxies, and black holes are believed to be responsible for the intervening phase transition known as the epoch of reionization \citep{LF13}. Through study of the reionization process, we hope to understand the nature of the faintest and earliest collapsed structures.

The CMB itself provides an ``integral" constraint on the total column density of ionized gas through the measured optical depth to electron scattering $\tau_e$, and thus a characteristic reionization redshift $z_\mathrm{reion}\sim8$ \citep{Planck16a}. However, a wide range of models for the reionization history are consistent with CMB measurements \citep{Planck16b}, and $\tau_e$ alone does not constrain the topology of the inhomogeneous reionization process \citep{Furlanetto04}. An important boundary condition to the epoch of reionization is the ubiquitous transmission through the \lya forest at $z\la5.5$ which is inconsistent with a predominantly neutral IGM \citep[e.g.][]{McGreer11,McGreer15}. Large stretches of neutral gas are still possible at higher redshift $z\ga6$ where the \lya forest is almost entirely opaque.

\begin{figure}
\resizebox{8.8cm}{!}{\includegraphics[trim={2em 0.5em 14em 0em},clip]{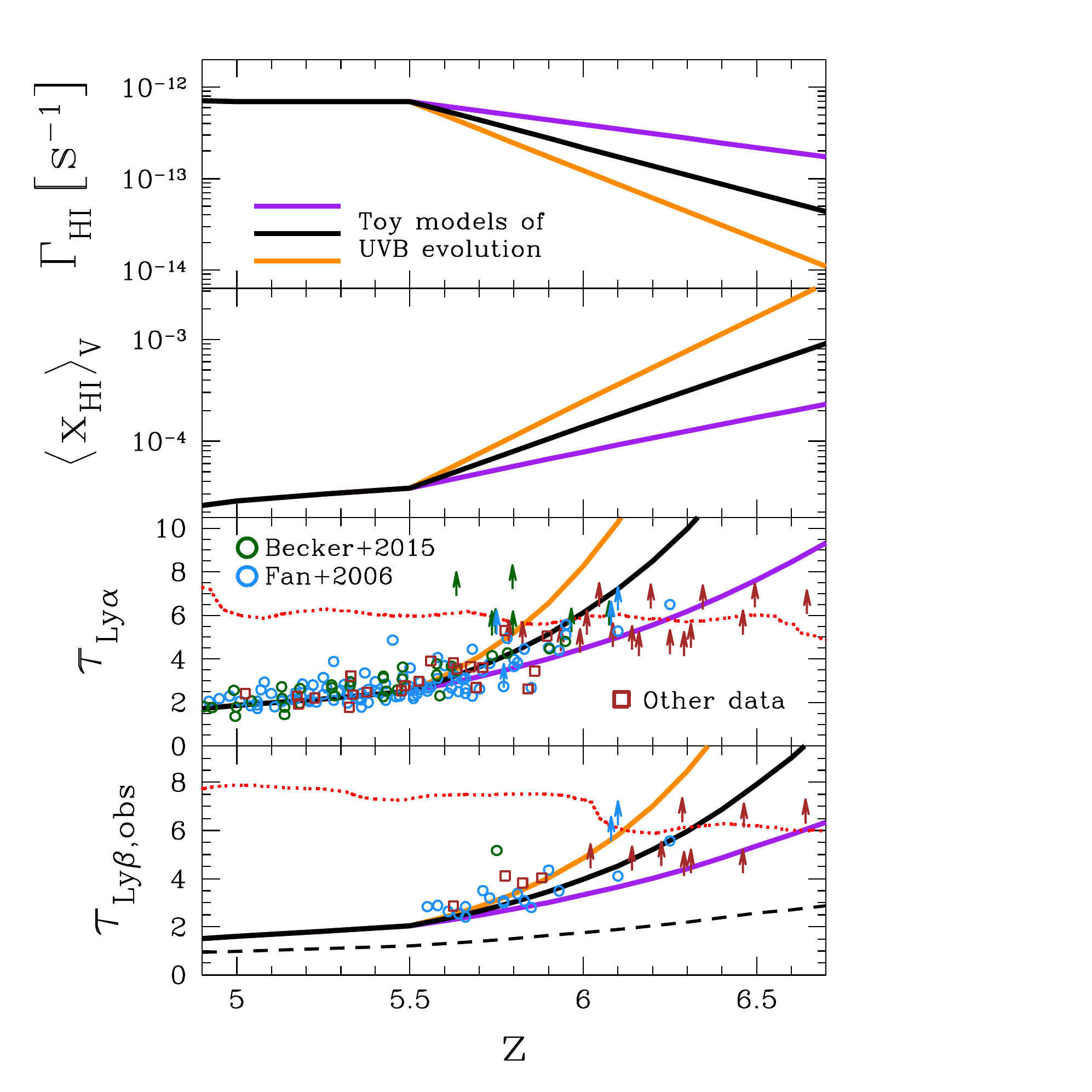}}
\caption{The top two panels show toy models for an evolving ionization rate $\GHI$ and the corresponding volume-weighted IGM neutral fraction $\langle x_\HI \rangle_V$. The bottom two panels show the 50 Mpc$/h$ \lya and \lyb forest $\teff$ measurements in \citet{Fan06} and \citet{Becker15} as the blue and green open circles, respectively, compared to simulated $\teff$ (solid curves) corresponding to the ionizing background evolution models in the top panel. The brown squares show quasar measurements not present in the other compilations \citep{Willott07,Mortlock09,Goto11,Tang17,Barnett17} and observations of high-redshift gamma ray bursts \citep{Chornock13,Chornock14}. Upward arrows represent 2$\sigma$ upper limits for measured transmission values that fall below reported 2$\sigma$ uncertainties. The dotted curve in the bottom two panels shows the 2$\sigma$ limiting effective optical depth on 50 Mpc$/h$ scales for a simulated 10 ks Keck spectrum of a bright high-redshift quasar ($F_{\mathrm{cont}}=10^{-17}$ erg s$^{-1}$ cm$^{-2}$ \AA$^{-1}$). The long-dashed curve in the bottom panel shows the contribution of the foreground \lya forest to the \lyb forest $\teff$.}
\label{fig:tauevol}
\end{figure}

The $z\sim6$ \lya forest has been studied for well over a decade to
constrain the evolution of the ionization state of the Universe close
to the reionization epoch
\citep[e.g.][]{Becker01,White03,Songaila04,Fan02,Fan06}. As shown by
the data points in the second panel from the bottom of Figure~\ref{fig:tauevol}, above
$z\sim5.5$ the typical opacity -- and the typical variation between
different lines of sight -- increases rapidly. Eventually, the only
transmission seen in the \lya forest is in narrow transmission spikes,
likely corresponding to small-scale regions of the universe with particularly low
density \citep[e.g.][]{OF05}.
The large-scale transmission measurements then simply reflect the number and strength
of these spikes that fall into wide bins in each spectrum, where the size is defined either in terms of redshift interval \citep[${\Delta}z\sim0.15$; e.g.][]{Fan06}
or by comoving distance \citep[50 Mpc$/h$;][]{Becker15}, both roughly corresponding to the scale of the first large-scale opaque regions, known as Gunn-Peterson troughs (after \citealt{GP65}; henceforth GP troughs), observed in the spectra of $z>6$ quasars \citep{Fan01,Becker01,Fan03,White03}.
Indeed, some large-scale regions with formal limits on their \emph{mean} transmission
contain high significance transmission spikes \citep{Becker15}, which
results in mean transmission below the significance threshold ($2\sigma$) when combined with surrounding GP troughs. Clearly, there is information in these small-scale transmission features which is lost when averaging on large scales, suggesting a ``matched filtering" approach targeting transmission spikes would be
more sensitive.

Another limiting factor for determining the ionization state of the $z\ga6$ Universe is that the \lya forest becomes essentially completely saturated, even at small scales, with many lines of sight consistent with zero transmission. The red dotted line in the bottom panels of Figure~\ref{fig:tauevol} shows the $2\sigma$ limiting optical depth for 10 ks exposure time on a bright ($F_\mathrm{cont} = 10^{-17}$ erg s$^{-1}$ cm$^{-2}$) high-redshift quasar with a 10-meter telescope. The solid curves of the panel second from the bottom show the evolution of the \lya forest opacity computed from the hydrodynamical simulation described in \S~\ref{sec:sims} assuming different toy models for the evolution in the hydrogen ionization rate $\GHI$ and corresponding evolution in the volume-averaged hydrogen neutral fraction $\langle x_{\rm HI} \rangle_V$ shown in the top panels. 

The \lyb forest saturates later because of its weaker oscillator strength, $f_\lybm \lambda_\lybm \sim 0.16 f_\lyam \lambda_\lyam$ where $f_i$ and $\lambda_i$ are the oscillator strength and rest-frame wavelength of transition $i$, respectively. In practice the interpretation of \lyb measurements is complicated by the presence of lower redshift \lya absorption, which at $z_{\lybm}\sim6$ corresponds to $z_{\lyam}\sim5$. Fortunately, the $z\sim5$ \lya forest has relatively small scatter and is well-understood compared to the strong fluctuations seen at $z>5.5$ \citep[e.g.][]{Becker15} and its statistics can be modeled with standard simulations of the IGM that employ a uniform ionizing background. The bottom panel of Figure~\ref{fig:tauevol} shows similar curves to the panel above it, but computed for the \lyb forest including the presence of foreground Ly$\alpha$, demonstrating that observable \lyb transmission should persist to substantially higher redshift than \lya at fixed observational cost. 

The most striking example of the increased sensitivity provided by \lyb is the $\sim110$ Mpc$/h$ GP trough observed in the $z=5.98$ quasar ULAS J0148+0600 with an effective optical depth $\teff>7.4$ ($2\sigma$ limit) in the \lya forest, but which shows many prominent transmission spikes in the \lyb forest (see Figure 5 in \citealt{Becker15}). The presence of \lyb transmission proves the existence of a highly-ionized IGM in this region despite the complete saturation of Ly$\alpha$, consistent with theoretical predictions for large-scale fluctuations in the ionizing background after the end of reionization \citep{DF16} or relic temperature fluctuations due to inhomogeneous heating during reionization (\citealt{D'Aloisio15}; Davies et al., in prep.).

In this work, we present theoretical expectations for the distribution of \lya and \lyb forest transmission on varying $\ga$Mpc scales at $z\sim6$, and demonstrate their ability to constrain the evolving ionizing background at high redshift. To account for sparseness, correlations, and extreme-value characteristics of potential data sets we apply a statistical method known as Approximate Bayesian Computation (ABC). ABC approximates the likelihood function through a massive Monte Carlo of mock data realizations,
in contrast to more typical methods that assume a functional form (e.g. a multivariate Gaussian distribution).
In \S~\ref{sec:sims}, we describe our numerical modeling of the \lya and \lyb forests and fluctuations in the ionizing background and present simulated distributions of \lya and \lyb forest transmission. In \S~\ref{sec:abc}, we describe our approach to measure the strength of the ionizing background using ABC.
In \S~\ref{sec:poc}, we apply our statistical method to real data as a proof-of-concept test, providing (model-dependent) constraints on $\GHI$ at $z\ga5.5$ from a single high-redshift quasar spectrum. In \S~\ref{sec:end}, we conclude with a summary and a discussion of predicted constraints from future samples of $z>6$ quasar spectra, where the addition of more quasars should both tighten the constraints on $\GHI$ and allow for constraints on the topology of ionizing background fluctuations.

All distance units are comoving unless specified otherwise. We assume a $\Lambda$CDM cosmology with
 $\Omega_m=0.3$, $\Omega_\Lambda=0.7$, $\Omega_b=0.047$, $h=0.685$, and $\sigma_8=0.8$.

\section{Modeling the \lya and \lyb forests}\label{sec:sims}
In this section, we describe our method for modeling the \lya and \lyb forests in the spectrum of $z\ga6$ quasars.
Our physical model for the IGM is a cosmological hydrodynamical simulation using the Nyx code \citep{Almgren13,Lukic15}, 100 Mpc$/h$ on a side, with 4096$^3$ dark matter particles and 4096$^3$ baryon fluid cells. We use outputs of the density, temperature, and velocity fields at $z=6.0$ and $z=5.0$ to model the $z\sim6$ \lya and \lyb forest and the $z\sim5$ \lya contamination of the $z\sim6$ \lyb forest, respectively. We extract random skewers from these simulation outputs, compute the \ion{H}{1} fraction in every cell assuming ionization equilibrium, and then calculate \lya and \lyb transmission along the line of sight including the effects of peculiar motions and thermal broadening (see \citealt{Lukic15} for more details). We re-scale the physical gas density along each skewer by $(1+z)^3$ to account for cosmological expansion, making the simplifying assumption that within the redshifts of interest the evolution of the overdensity field is less important.

Our simulation meets the suggested mass resolution requirement to resolve the $z=5$ \lya forest proposed by \citet{BB09} ($M_\mathrm{gas}\la3\times10^5$ M$_\odot$), although convergence may be different for a grid code like Nyx versus the smoothed-particle hydrodynamics (SPH) simulations they studied. Convergence may also be more difficult for the extremely low transmission of the $z\ga6$ \lya forest, suggested by the lack of convergence between the co-spatial \lya and \lyb optical depths at $z\sim6$ of relatively high-resolution simulations shown in the Appendix of \citet{Becker15}. In future work we will quantify the convergence (or lack thereof) of the $z\ga6$ IGM in our simulation, extending the tests in \citet{Lukic15} which were limited to $z\leq4$, but preliminary tests suggest that the mean \lya transmission is converged to better than $\sim10\%$ (J. O\~{n}orbe, private communication).
We note that the convergence of the $z=5$--$6$ \lya forest is assisted by the non-adaptive spatial resolution of the Eulerian grid in the most underdense environments which lead to \lya forest transmission \citep{OF05} (as opposed to, e.g., nearest-neighbor smoothing kernels in SPH codes).

\begin{figure}
\begin{center}
\resizebox{8cm}{!}{\includegraphics[bb=0 0 350 350,trim={2em 3em 6em 0em}]{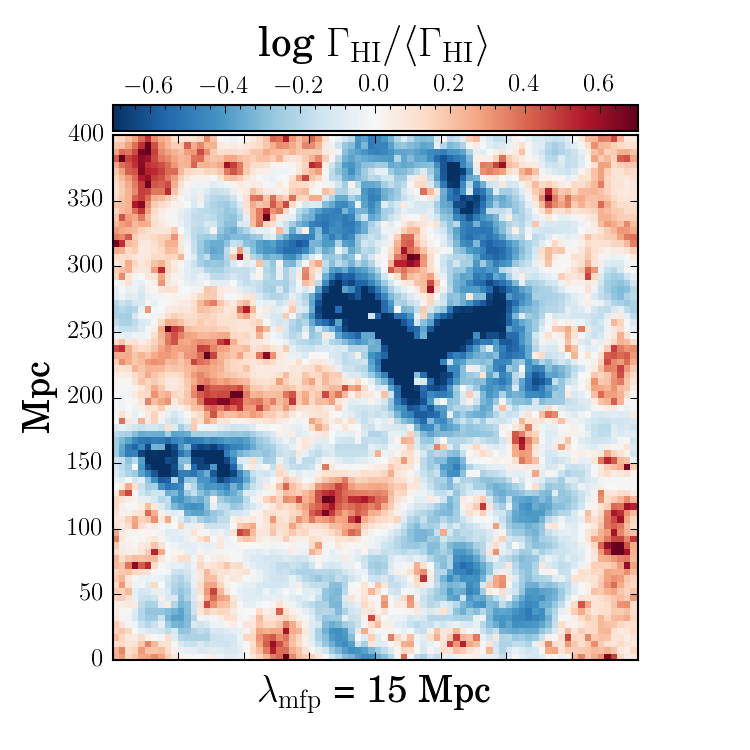}}
\end{center}
\caption{Slice through our simulation of the fluctuating ionizing background at $z=6$ computed as in \citet{DF16} with the mean free path at the average strength of the ionizing background fixed to 15 Mpc. Strong large-scale fluctuations exist on $\sim100$ Mpc scales as required by observations of Gunn-Peterson troughs at $z\ga5.5$ \citep{Fan06,Becker15}.}
\label{fig:uvb}
\end{figure}

\begin{figure*}
\begin{center}
\resizebox{18.5cm}{!}{\includegraphics[trim={6em 35em 0 0},clip]{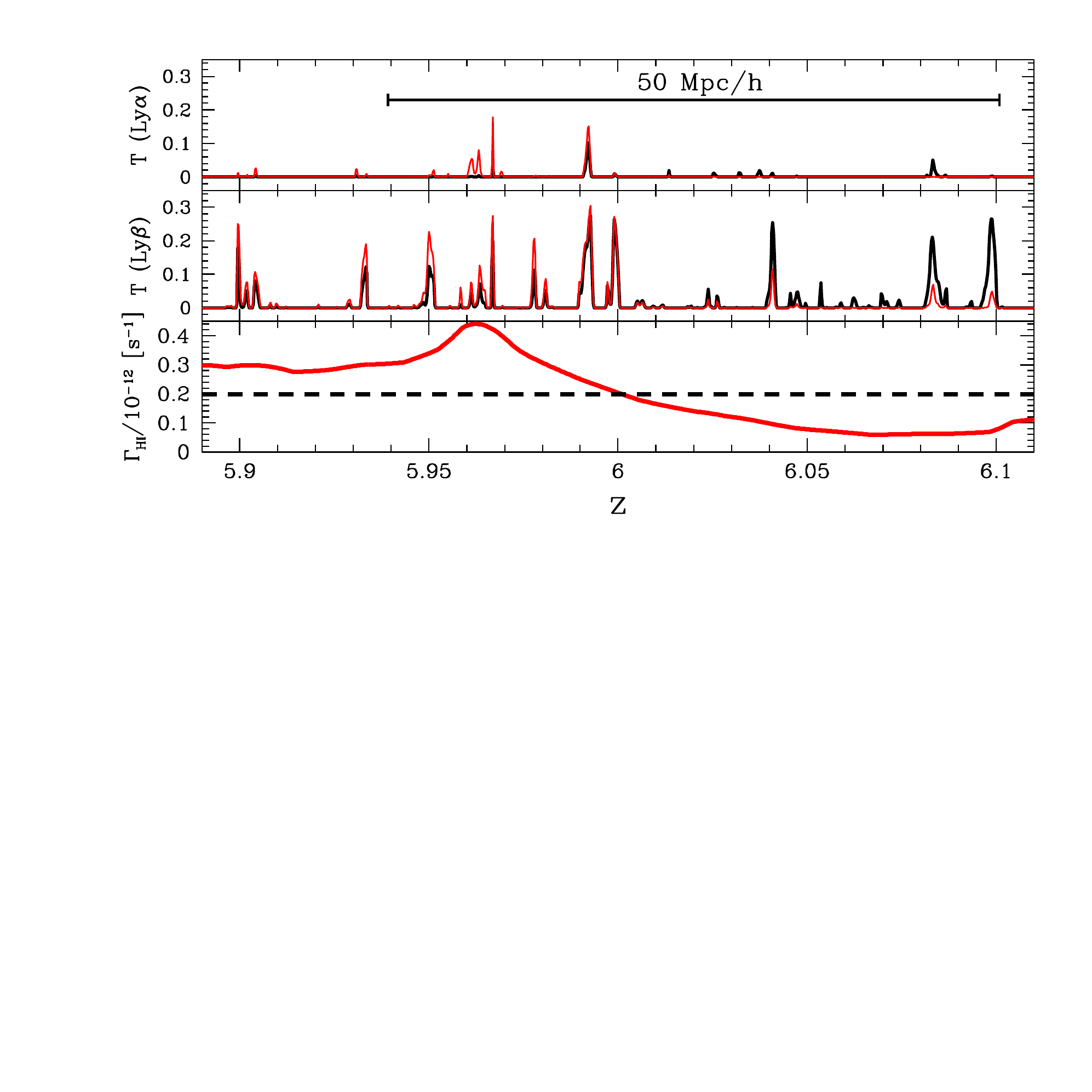}}
\end{center}
\caption{The top and middle panels show mock (noiseless) \lya and \lyb forest spectra, respectively, at $z\sim6$. The black curve assumes a uniform ionization rate $\GHI=10^{-12.7}$ s$^{-1}$, the dashed curve in the bottom panel, while the red curve demonstrates the effect of a fluctuating ionizing background skewer shown by the solid red curve in the bottom panel. In the upper panel we show a bar corresponding to the typical scale of mean transmission measurements (50 Mpc$/h$) in the \lya forest from the literature \citep[e.g.][]{Fan06,Becker15}.}
\label{fig:mockspec}
\end{figure*}

We include noise in our mock spectra using a noise vector extracted from a spectrum of the $z\sim6.4$ quasar SDSS J1148+5251 taken with the Echellete Spectrograph and Imager \citep[ESI;][]{Sheinis02} on the Keck II telescope
 with a total exposure time of 13,200s (PI: Cowie), re-scaled to an effective exposure time of 10,000s by multiplying the noise by $\sqrt{13200/10000}\sim1.15$. The noise vector includes contributions from read noise, sky noise, and the varying sensitivity of ESI as a function of wavelength. We assume that the noise is independent and Gaussian-distributed in each pixel and that the presence of source signal does not affect the noise level, a good assumption for the faint $z\ga6$ quasars for which the sky background dominates the noise budget. The quasar spectrum is assumed to follow the composite spectrum of \citet{Lusso15} assuming a flux density at 1450{\AA} of $F_\lambda(1450{\rm \AA}) = 10^{-17}$ erg s$^{-1}$ cm$^{-2}$ \AA$^{-1}$, corresponding to a fairly bright $z\sim6$ quasar with $M_{1450}\sim-27$.

We model fluctuations in the post-reionization ionizing background in
a separate (400 Mpc)$^3$ cosmological
volume using the method of \citet[][henceforth DF16]{DF16}. We defer the interested reader to DF16 for the details, but we briefly summarize the method here. We use the semi-numerical simulation code {\small DEXM} \citep{MF07} to generate a realization of cosmological initial conditions in a (400 Mpc)$^3$ volume and populate it with dark matter halos down to $M_\mathrm{h,min}=2\times10^9\ M_\odot$ following the excursion set formalism. We then abundance match these halos to the measured Bouwens et al. (2015) $z\sim6$ rest-frame UV luminosity function of galaxies down to absolute magnitudes of $M_\mathrm{UV,max}\sim-12.5$, and assume a constant ratio between the non-ionizing UV luminosity and the (escaping) ionizing luminosity of each galaxy. The ionizing background from these galaxies is computed on a coarse 80$^3$ grid allowing the mean free path of ionizing photons to vary spatially depending on the large-scale density field and local ionizing background, $\lambda \propto \GHI^{2/3}\Delta^{-1}$, with the average mean free path in the volume normalized to $\lambda_\mathrm{mfp}=15$ Mpc. In Figure~\ref{fig:uvb} we show a slice through the ionizing background model, which exhibits strong large-scale fluctuations consistent with what is required to reproduce the distribution of Gunn-Peterson troughs in the $z\ga5.5$ \lya forest (DF16). 

We simulate the ionizing background in a separate volume from the hydrodynamical simulation because the latter is not large enough to fully model the effect of the fluctuating mean free path of ionizing photons, and thus would have much weaker ionizing background fluctuations on the $\sim100$ Mpc scales we investigate in this work (DF16). Because of this, we are explicitly ignoring the anti-correlation of the density and radiation fields on very large scales shown by DF16. In this work we assume that this large-scale anti-correlation does not strongly affect the statistics of small-scale transmission features in the IGM, an assumption we will test in future work.

\setcounter{footnote}{0} 

In the top and middle panels of Figure~\ref{fig:mockspec}, we show \lya and \lyb forest segments of a mock spectrum from $z=5.9$--$6.1$. As expected, the \lyb forest shows considerably more transmission, and despite contamination by
foreground $z\sim5$ \lya forest absorption, transmission features in the \lya forest tend to show counterparts in the \lyb forest. The red curves show the effect of applying the fluctuating ionizing background skewer in the bottom panel. Small-scale features in the \lya forest are sensitive to ionizing background fluctuations because the optical depth in each pixel changes as $\tau_\alpha \propto \GHI^{-1}$. The \lyb forest is less sensitive to these fluctuations
because $\tau_\beta$ contains a constant factor from the foreground \lya forest contamination which sets a floor to the opacity, i.e. $\tau_\beta = \tau_{\beta,z\sim6} + \tau_{\alpha,z\sim5}$ where only the first term responds to $z\sim6$ background fluctuations\footnote{In this work we assume that the ionizing background at $z\sim5$ is uniform, a reasonable assumption given the relatively large measured mean free path of ionizing photons (\citealt{Worseck14}, but see \citealt{D'Aloisio17}) and the agreement between the distribution of large-scale \lya forest optical depth with standard hydrodynamical simulations \citep{Becker15}.}.

\subsection{\lya and \lyb Transmission PDF on 2 \& 20 Mpc Scales}\label{sec:pdfs}
\begin{figure*}
\begin{center}
\resizebox{18.5cm}{!}{\includegraphics[trim={0 30em 0 0},clip]{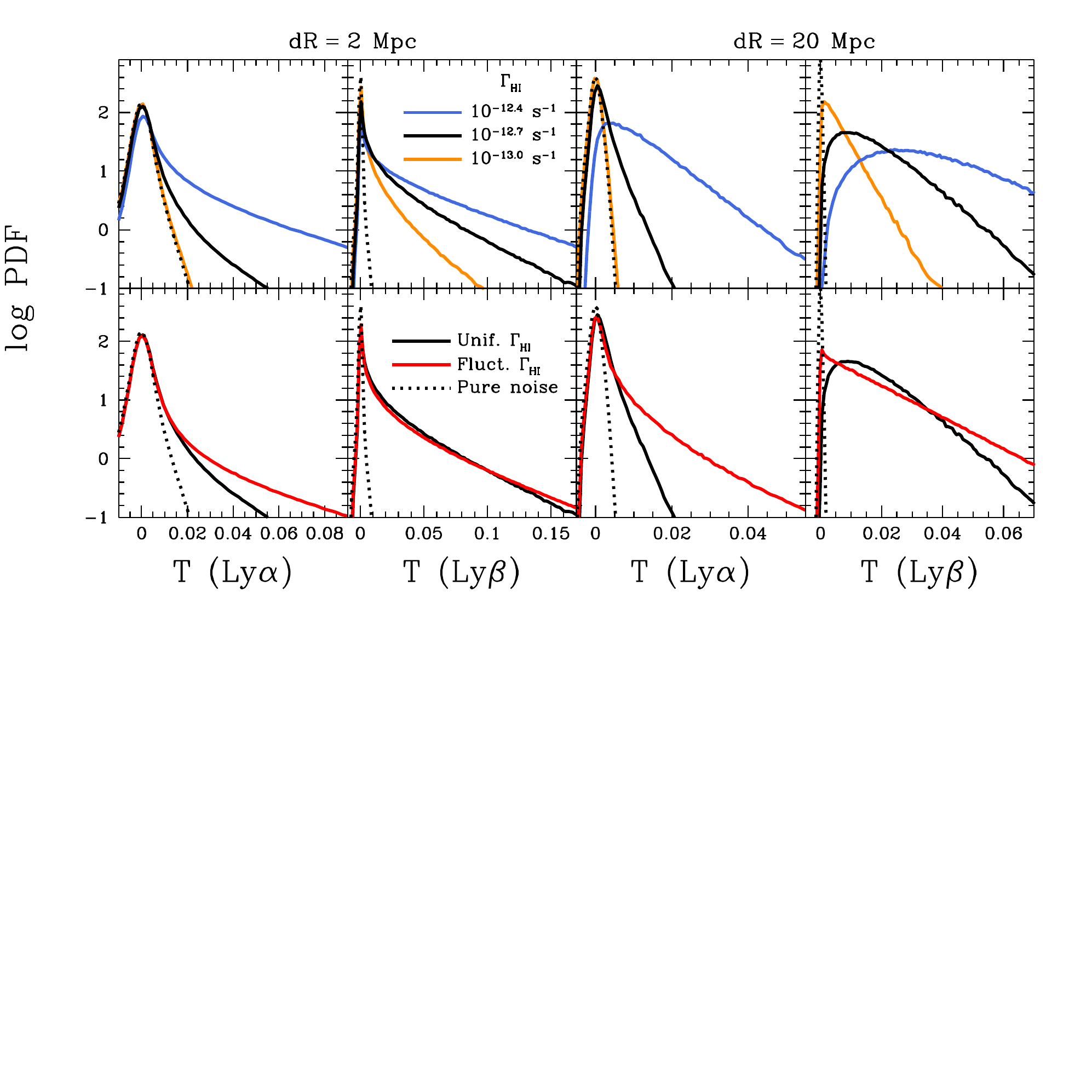}}
\end{center}
\caption{Simulated PDFs of 2 Mpc (left four panels) and 20 Mpc (right four panels) binned segments of the \lya (left panels inside each $dR$) and \lyb (right panels inside each $dR$) forests at $z\sim6$. The upper panels show the variation in the PDFs as a function of the (uniform) ionization rate $\GHI=10^{-13.0},$ $10^{-12.7}$, and $10^{-12.4}$ s$^{-1}$ as the orange, black, and blue curves, respectively. The lower panels show the effect of including fluctuations in the ionizing background (red curve) with an average $\GHI=10^{-12.7}$ s$^{-1}$. The dotted curve in every panel shows the PDF of the noise model, which assumes independent Gaussian-distributed noise according to a noise vector from a real Keck/ESI spectrum.}
\label{fig:pdfs}
\end{figure*}

In the upper left set of panels in Figure~\ref{fig:pdfs} we show the transmission probability distribution function (PDF) on 2 Mpc scales (d$z\sim0.005$) of the \lya and \lyb forests from simulations of $z\sim6$ spectra with varying uniform $\GHI$ of $10^{-12.4}$ s$^{-1}$ (top, blue), $10^{-12.7}$ s$^{-1}$ (middle, black), and $10^{-13.0}$ s$^{-1}$ (bottom, orange). 
These $\GHI$ values roughly correspond to the mean and +/- 1$\sigma$ uncertainty range of current measurements from the \lya forest at $z\sim6$ \citep{WB11}.
 The PDFs are characterized by a Gaussian peak around zero -- corresponding to the noise
floor, shown as the dotted line in every panel -- and a tail towards higher transmission whose steepness is a function of ionizing background strength. The shape of the PDF is markedly different from the distributions seen at $z\sim2$--$3$ \citep[e.g.][]{Lee15} because
the $z\ga6$ \lya and \lyb
forests are so opaque that they almost never come close to the continuum \citep[see also][]{Becker07}. In the bottom panels the red curve corresponds to a model with the same average $\GHI=10^{-12.7}$ s$^{-1}$ that includes fluctuations in the ionizing background (as shown in Figures~\ref{fig:uvb} and \ref{fig:mockspec}).
In general the effect of background fluctuations on the 2 Mpc-binned
transmission PDF is minor, although it does lead to an extended tail of rare transmission spikes.
The effect is larger on the \lya PDF than the \lyb PDF due to the contamination of \lyb by foreground $z\sim5$ \lya forest absorption, i.e., even if $\GHI$ fluctuates high, the transmission spikes can only increase up to the absorption level of the $z\sim5$ \lya forest (\S~\ref{sec:sims}).

In the right set of panels in Figure~\ref{fig:pdfs} we show the PDFs on 20 Mpc scales. This is analogous to previous studies of large bins in the \lya forest \citep{Fan06,Becker15} but at a scale smaller by around a factor of three. The distributions are narrower (note the difference in transmission scale on the horizontal axis), reflecting not only a decrease in sky noise (S/N$\,\propto N_\mathrm{pix}^{-1/2}$), but also the tendency for several transmission spikes to fall into a single bin leading to a somewhat more Gaussian shape to the distribution (as per the central limit
theorem). For moderate
ionization rates $\GHI\ga10^{-12.7}$ s$^{-1}$, the model PDFs suggest
that high signal-to-noise transmission should nearly always be
measured in the \lyb forest at this scale. The effect of ionizing
background fluctuations on the PDFs is considerably stronger on large
scales, consistent with the $\ga10$s of Mpc typical scales of
background fluctuations seen in the simulations (e.g. Figure~\ref{fig:mockspec}).
In
particular the large-scale \lyb PDF is strongly affected by ionizing
background fluctuations through an excess in 20 Mpc-scale \lyb-dark
regions (GP troughs) over the uniform background model, which are otherwise consistent with the noise floor in the \lya forest, suggesting that it may have
considerable constraining power for the parameters of models for
strong fluctuations in IGM opacity
(e.g. DF16, \citealt{D'Aloisio15}).
The 20 Mpc-scale PDFs shown here are analogous to the 50 Mpc$/h$-scale (cumulative) PDFs studied by \citet{Becker15} -- 
they similarly allow one to characterize fluctuations in the \lya and \lyb forests on the large-scales expected from an inhomogeneous reionization process \citep[e.g.][]{Furlanetto04}.

From these PDFs we should be able to assess the probability for any given model to reproduce rare transmission spikes in the \lya and \lyb forests, and from their detailed shape, study the patchy nature of the post-reionization IGM. In practice, performing parameter inference from the PDFs is difficult because much of their constraining power lies in a handful of transmission spikes in the tail of the distribution, and the features in the \lya and \lyb forests should be correlated because they trace the same physical structures.

\section{Statistical Methods: Approximate Bayesian Computation}\label{sec:abc}

The standard approach to IGM parameter inference from the transmission PDF of the \lya forest at $z\sim2$--$3$ \citep[e.g.][]{Lidz06} is to choose a set of coarse bins in the range of $0 < F < 1$, compute the covariance matrix of the bins from either jackknife sampling of real data \citep{Lidz06,Lee15} or forward-modeling of the \lya forest in hydrodynamical simulations \citep{Rollinde13}, and then approximate the likelihood function as a multivariate Gaussian. Because each PDF bin typically contains thousands of pixels, the central limit theorem ensures that a multivariate Gaussian likelihood is a good approximation.

\begin{figure}
\epsscale{1.2}
\plotone{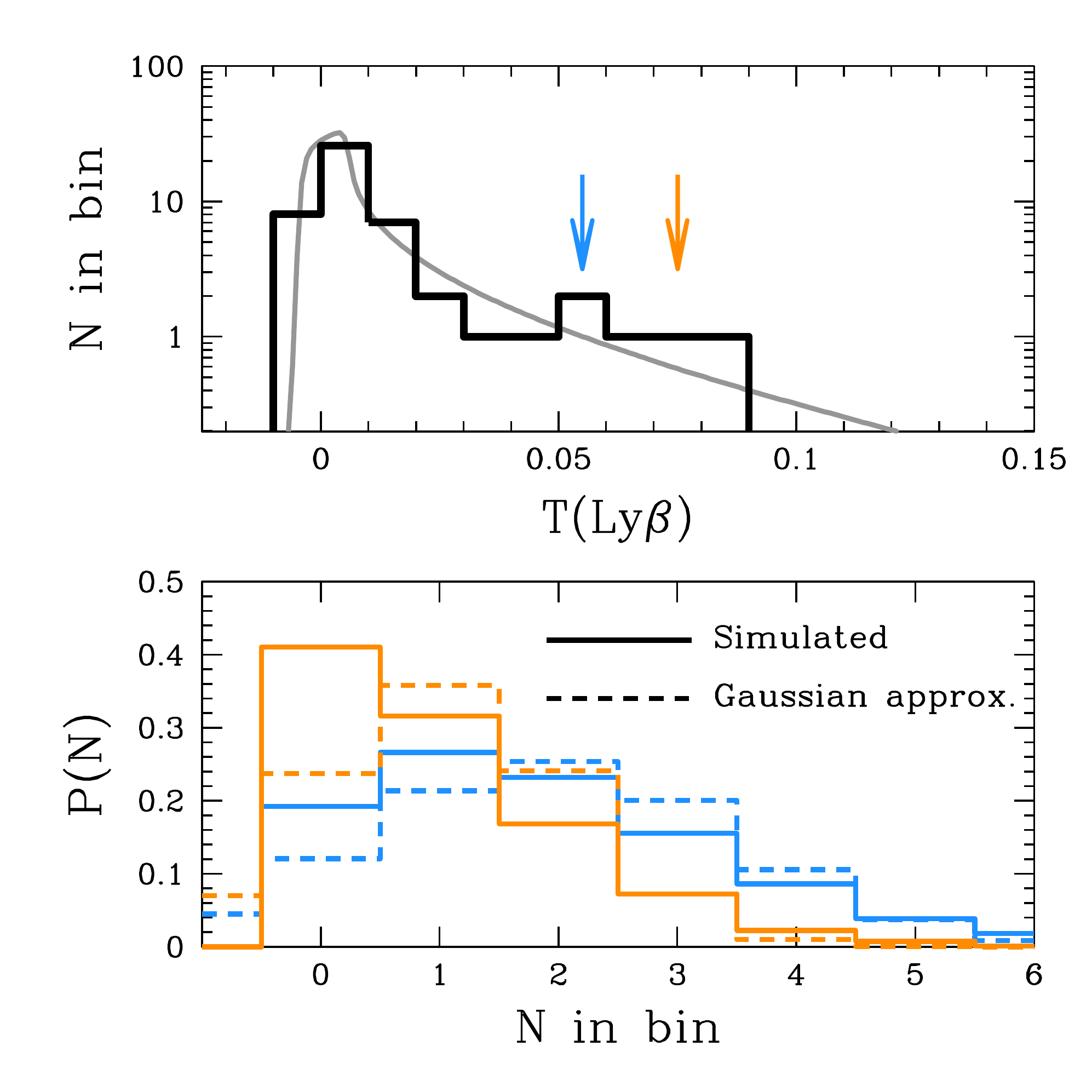}
\caption{Top: The histogram shows the $\Delta T=0.01$-binned transmission distribution of a single mock observation of the \lyb forest covering 100 Mpc with transmission binned on 2 Mpc scales, with $\GHI=10^{-12.7}$ s$^{-1}$ and \mfp$=15$ Mpc at $z\sim5.9$--$6.1$. The curve shows the model PDF convolved with a $\Delta T=0.01$ boxcar filter to be consistent with the binned PDF. Bottom: The solid and dashed histograms represent the true and Gaussian approximations to the distribution of PDF values, respectively. The two sets of histograms represent the two bins labeled (and color-coded) by the arrows in the top panel.}
\label{fig:nongauss}
\end{figure}

The situation changes dramatically at higher redshifts where the IGM is more opaque and fewer quasars are available. At $z\ga6$, most pixels in the \lya and \lyb forests are consistent with zero transmission, with only a \emph{sparse} tail of real transmission spikes. While the (co)variances of each bin can be estimated by forward modeling of mock spectra, the resulting Gaussian distributions (as defined by the mean and (co)variance of the mock samples) are not representative of the distributions of the number of pixels that fall into each bin. We show an example of this in Figure~\ref{fig:nongauss}, where the bottom panel illustrates the true and Gaussian-approximated distributions of mock \lyb PDF measurements in transmission bins highlighted in the top panel, with one mock observation shown. 

The requirement of approximately Gaussian likelihood is more stringent for transmission measurements on larger bin scales, such as the $50$ Mpc$/h$ measurements by \citet{Becker15}, where the small number of measurements -- and the tendency for measurements to be reported as limits -- have typically led to rough ``chi-by-eye" comparisons to simulations \citep[e.g.][]{D'Aloisio15,DF16,D'Aloisio17}.

To circumvent these challenges, we adopt the statistical approach known as
ABC (e.g. \citealt{Pritchard99}; see also \citealt{CP12,Ishida15,Hahn16,Jennings16} for other astronomical applications) which allows for parameter inference without any explicit analytic form for the likelihood function.
We start with an abridged statement of
Bayes' Theorem,
\begin{equation}
p(\theta|{\bf d}) \propto p({\bf d}|\theta)p(\theta),
\end{equation}
where $\theta$ represents the set of model parameters (e.g. $\GHI$) and ${\bf d}$ represents the data (e.g. the transmission measured in each spectral pixel), $p({\bf d}|\theta)$ is the likelihood function, $p(\theta)$ is the prior distribution representing our previous knowledge of the model parameters, and $p(\theta|{\bf d})$ is the posterior distribution for $\theta$ that we hope to obtain. In many cases, the likelihood function can be approximated as a multivariate Gaussian distribution or can be straightforwardly computed numerically,
allowing the posterior distribution to be computed through methods such as Markov Chain Monte Carlo (MCMC). In our case, as we have shown above, such an approximation fails to adequately describe the distribution of mock measurements in our models.

One could imagine a brute force computation of the likelihood function by computing mock data sets to determine the probability of exactly reproducing the data, but this is computationally intractable for many problems.
The first and most fundamental approximation proposed by ABC is to slightly blur the definition of the likelihood function, instead describing the posterior distribution
as \citep[e.g.][]{Marin12}
\begin{equation}
p(\theta|{\bf d}) \approx p_\mathrm{ABC}(\theta|{\bf d}) \propto \int p(\rho({\bf d},{\bf x}) < \epsilon, {\bf x}|\theta)p(\theta) d{\bf x},
\end{equation}
where ${\bf x}$ is a forward-modeled mock data set, $\rho({\bf d},{\bf x})$ represents a distance measure
between the data and mock data, $\epsilon$ is a small tolerance threshold, and the integral is performed over the space of forward-modeled mock datasets.
That is, to approximate the likelihood function, ABC proposes that coming \emph{close} to the measured data is ``good enough," allowing for far easier computation than the brute force method mentioned above. In the process, the correlations between data elements are naturally accounted for through the forward-modeling procedure without any explicit computation of the covariance matrix.

The simplest application of ABC in practice is to draw model parameter samples $\theta^*$ from the prior, forward model a mock realization of the data given by those parameters ${\bf x}^*$, then measure the distance between the mock data and real data $\rho({\bf d},{\bf x}^*)$. If this distance is ``small enough", i.e. $\rho({\bf d},{\bf x}^*) < \epsilon$, then $\theta^*$ is accepted as a sample of the posterior distribution. This procedure is called ABC rejection sampling.

The above methodology may still be computationally intractable if the data represent a highly stochastic process or high-dimensional data set. For example, in our case, producing a mock data set that nearly matches an observed \lya and/or \lyb forest spectrum at the pixel-to-pixel level (i.e. with small $\rho({\bf d},{\bf x})$) requires nearly matching the peak heights and positions of every narrow transmission feature. In practice this would require matching not only the ionization state of the gas, but also the \emph{entire set of cosmological initial conditions} that led to the
structures observed along the sightline.
To overcome this computational hurdle, the technique commonly employed in ABC is to use ``summary statistics" to process the data (and mock data) into a form which allows them to be compared more easily, typically into a lower-dimensional state. This reformulation can be stated as
\begin{eqnarray}
&&p(\theta|{\bf d}) \approx p(\theta|{\bf s}_\mathrm{obs}) \nonumber \\ && \approx p_\mathrm{ABC}(\theta|{\bf s}_\mathrm{obs}) \propto \int p(\rho({\bf s}_\mathrm{obs},{\bf s}) < \epsilon, {\bf x}|\theta)p(\theta) d{\bf x},
\end{eqnarray}
where ${\bf s}_\mathrm{obs}$ is the summary statistic of the data and ${\bf s}$ is the summary statistic of mock data set ${\bf x}$. If the summary statistic is ``sufficient" then no information is lost, but such summary statistics are only known analytically in very simple cases (e.g., for determining the mean of a normally-distributed quantity, the sample mean is a sufficient statistic). Thus, in most cases, the use of a summary statistic implies loss of information, leading to bias and/or loss of precision in the posterior distribution. In \S~\ref{sec:inftest}, we perform inference tests on multiple summary statistics of the \lya and \lyb forests to investigate their potential biases.

While ABC may seem like a ``silver bullet" to perform parameter inference with the least number of assumptions about the nature of the likelihood function, its primary weakness is the requirement of simulating an overwhelmingly large number of mock data sets in order to produce a reasonably converged approximation to the posterior PDF. Even with a one-dimensional summary statistic one may have to generate \emph{millions} of mock data sets, limiting its potential application for practical reasons. Nevertheless, ABC allows for inference involving complicated likelihood functions, fully accounting for covariance (depending on the sufficiency of the chosen summary statistic) \emph{implicitly} as they manifest in the assumed model for mock data.

\subsection{Application of ABC to the \lya and \lyb forests}\label{sec:abcapp}

Application of ABC to inference of the \lya and \lyb forests is performed as follows:

\begin{enumerate}
\item Draw a value of $\GHI$ from prior distribution, $\GHI^*$.
\item Compute mock spectrum of \lya and \lyb forest with $\GHI=\GHI^*$, including noise consistent with observations.
\item Compute summary statistic of mock and observed spectra (${\bf s}$ and ${\bf s}_\mathrm{obs}$, respectively).
\item Compute distance between ${\bf s}_\mathrm{obs}$ and ${\bf s}$, $\rho({\bf s}_\mathrm{obs},{\bf s})$.
\item If $\rho({\bf s}_\mathrm{obs},{\bf s})$ is below the threshold $\epsilon$, keep $\GHI^*$ as a sample of the \emph{approximate} posterior PDF, $p_\mathrm{ABC}(\GHI)$.
\item Repeat steps 1--5 until a predetermined number of posterior PDF samples have been drawn, e.g. 1000.
\end{enumerate}

In practice, the $\GHI^*$ can be drawn more efficiently, and in later sections we adopt a method based on importance sampling with an iteratively declining threshold $\epsilon$ set based on percentiles of the distribution of $\rho({\bf s}_\mathrm{obs},{\bf s})$ (\citealt{Beaumont09}; see detailed implementations by e.g. \citealt{Ishida15,Hahn16,Jennings16}). These procedures are simply methods to increase the efficiency of the basic rejection sampling method, and as such we will not discuss them in further detail.

For a given segment of the \lya or \lyb forest, perhaps the most basic summary statistic is the mean transmission: ${\langle}F{\rangle} = \sum_i^{N_\mathrm{pix}}F_i/N_\mathrm{pix}$. The use of this statistic reduces the dimensionality of the data from [$\#$ of \lya forest pixels + $\#$ of \lyb forest pixels]
to 2: the mean transmission in \lya and the mean transmission in Ly$\beta$. That is, ${\bf s} = (\langle F_\lyam \rangle, \langle F_\lybm \rangle)$. Using ABC, we do not have to make any assumptions about the nature of the intrinsic distributions of \lya and \lyb mean transmission, nor the details of their correlations. Instead, the process of comparison to forward modeled mock data naturally accounts for correlations under the assumption that our model for the IGM and the noise model of the spectrum is accurate.

\begin{figure}
\epsscale{1.2}
\plotone{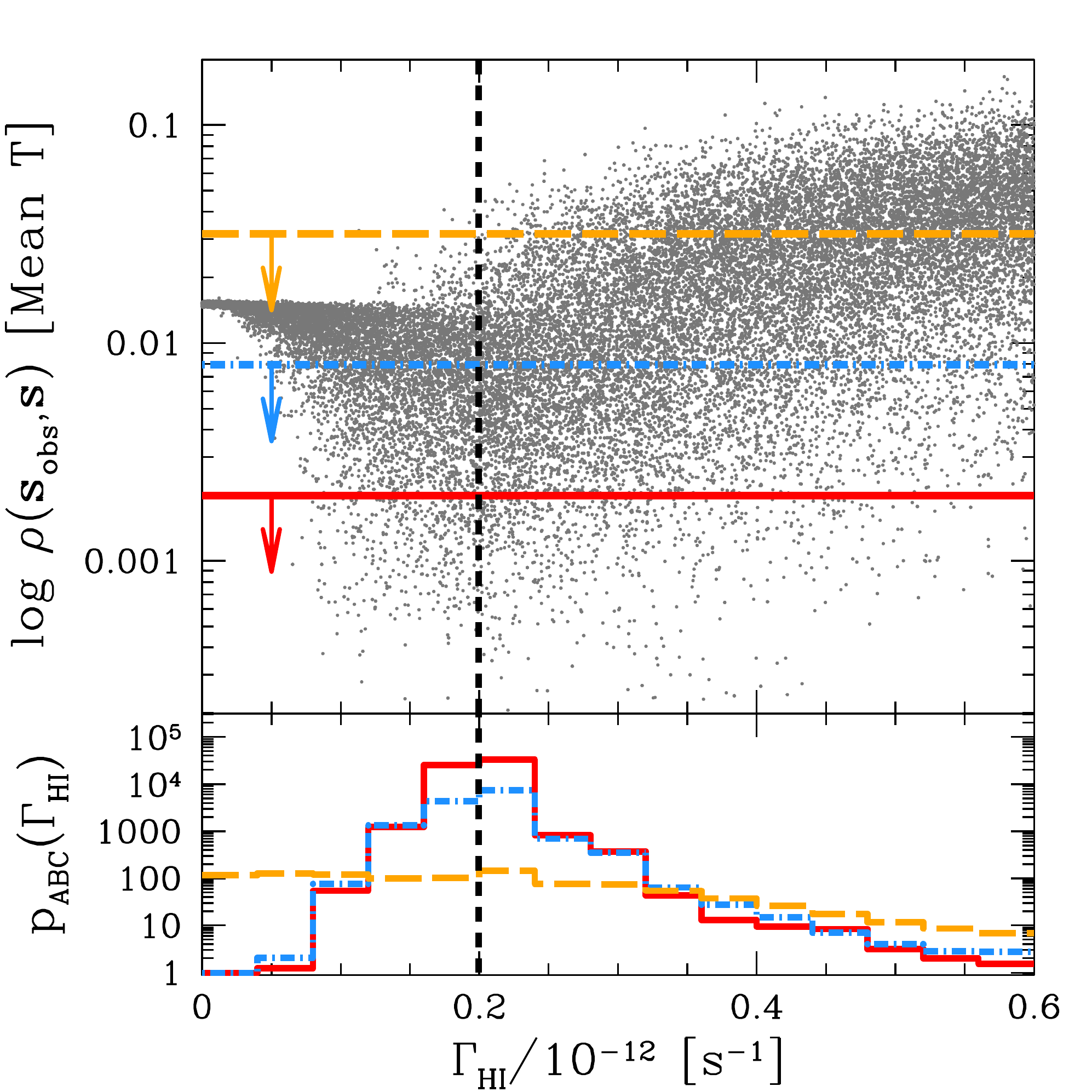}
\caption{Demonstration of ABC applied to a mock observation of the \lya and \lyb forests. The ``observation" consists of the mean transmissions of \lya and \lyb in a single mock spectrum covering 100 Mpc at $z\sim6$ with a fluctuating ionizing background and $\GHI=10^{-12.7}$ s$^{-1}$. The points in the upper panel represent the distances between the ``observed" mock spectrum and 25000 additional mock spectra with $\GHI$ drawn from a uniform prior, where the distance has been defined using the mean transmission as the summary statistic (equation~\ref{eqn:meansum}).  The thick dashed orange, dot-dashed blue, and solid red histograms in the lower panel represent the ABC posterior distributions assuming thresholds of log $\epsilon=$ -1.5, -2.1, and -2.7, respectively, corresponding to the horizontal lines in the upper panel.}
\label{fig:abcdemo}
\end{figure}

We set up an example of the ABC procedure as follows. One mock data set with known $\GHI$, specifically a noisy realization of the \lya and \lyb forest with fluctuating $\GHI$ shown in
Figure~\ref{fig:mockspec}, covering a path length of 100 Mpc,
is chosen to be the ``observed" sample. We then draw $\GHI^*$ from a uniform prior covering 0--$0.6\times10^{-12}$ s$^{-1}$, forward model a mock data set using that $\GHI^*$, then compute the distance between the observed and mock data sets,
\begin{eqnarray}\label{eqn:meansum}
&&\rho({\bf s}_\mathrm{obs},{\bf s}) = \nonumber \\ && \sqrt{(\langle{F_\mathrm{obs}^\alpha}\rangle-\langle{F_\mathrm{mock}^\alpha}\rangle)^2+(\langle{F_\mathrm{obs}^\beta}\rangle-\langle{F_\mathrm{mock}^\beta}\rangle)^2},
\end{eqnarray}
where we have adopted the $L_2$ norm for the distance $\rho$.

The result of such a test with 25000 mock data sets is shown in Figure~\ref{fig:abcdemo}.
Each point in the top panel represents the distance defined above for a single mock spectrum corresponding to a draw of $\GHI^*$, a random IGM skewer from the hydrodynamical simulation, and a random fluctuating ionizing background skewer. The bottom panel shows binned samples from the posterior distribution of $\GHI$ obtained by rejecting samples with distance greater than the $\epsilon$ values given by the horizontal lines in the upper panel (i.e. the samples below the lines are accepted). As the tolerance $\epsilon$ is decreased, the ABC posterior distribution narrows, representing successive improvements to the approximated likelihood as it converges to the true distribution.
Through the ABC process, this posterior distribution explicitly includes cosmic variance (as modeled by the cosmological simulation), measurement uncertainty, and the correlation between the \lya and \lyb forest mean transmissions.

\begin{figure}
\epsscale{1.1}
\plotone{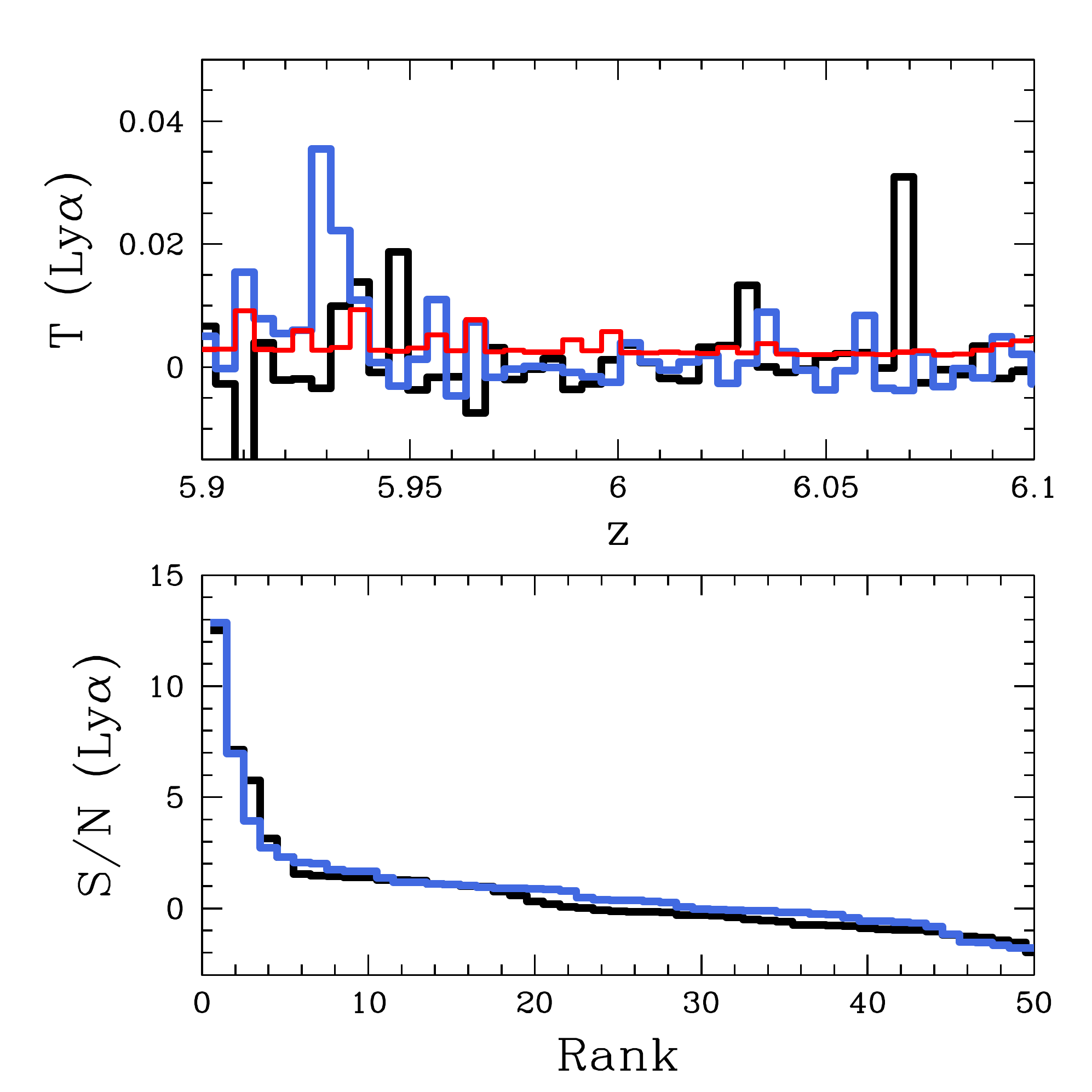}
\caption{Top: Two mock \lya forest spectra (black, blue) at $z\sim6$ computed with the same ionization rate $\GHI=10^{-12.6}$ s$^{-1}$ and binned to 2 Mpc. The noise vector is shown as the red curve. Bottom: Rank-order distribution of the 2 Mpc binned signal-to-noise of the spectra shown in the upper panel. The ``distance" between the two spectra, qualitatively the amount of ``space" between the black and blue curves, is much smaller when arranged in rank order.}
\label{fig:rankdemo}
\end{figure}

However, the mean transmissions in the \lya and \lyb forests
clearly do not tell the entire story. Due to sky noise, a large region of forest may have a mean transmission consistent with zero but still contain a significant positive spike (e.g. Figure 2 in \citealt{Becker15}). The presence of any (real) transmission spike requires an ionizing background greater than zero, but the mean transmission alone may not contain this information because of limited signal-to-noise.
A summary statistic which retains this information but disregards the locations of transmission spikes
is the \emph{rank-order distribution} of transmission on small scales. That is, a re-ordering of the transmission values such that they are in order from lowest to highest transmission. The scale over which the transmission is binned
can be chosen to maximize the signal-to-noise ratio of transmission spikes, i.e. matched filtering. We also account for the spatially-varying noise of the spectrum (e.g. due to sky lines) by first dividing each transmission value by the local noise before re-ordering the measurements. An example of rank-order distributions of two mock \lya forest spectra with the same $\GHI$ is shown in Figure~\ref{fig:rankdemo}. The distance between observed and mock data is then
\begin{eqnarray}
&&\rho({\bf s}_\mathrm{obs},{\bf s}) = \nonumber \\ && \sqrt{\sum_i^{N_\alpha} (S_{i,\mathrm{obs}}^\alpha-S_{i,\mathrm{mock}}^\alpha)^2 + \sum_i^{N_\beta} (S_{i,\mathrm{obs}}^\beta-S_{i,\mathrm{mock}}^\beta)^2},
\end{eqnarray}
where $S_i$ is the signal-to-noise ratio of the transmission in the bin with rank $i$ (where the \lya and \lyb forests are ranked independently), and $N_{\alpha}$ and $N_{\beta}$ are the number of bins in the \lya and \lyb forests, respectively.

\begin{figure}
\epsscale{1.1}
\plotone{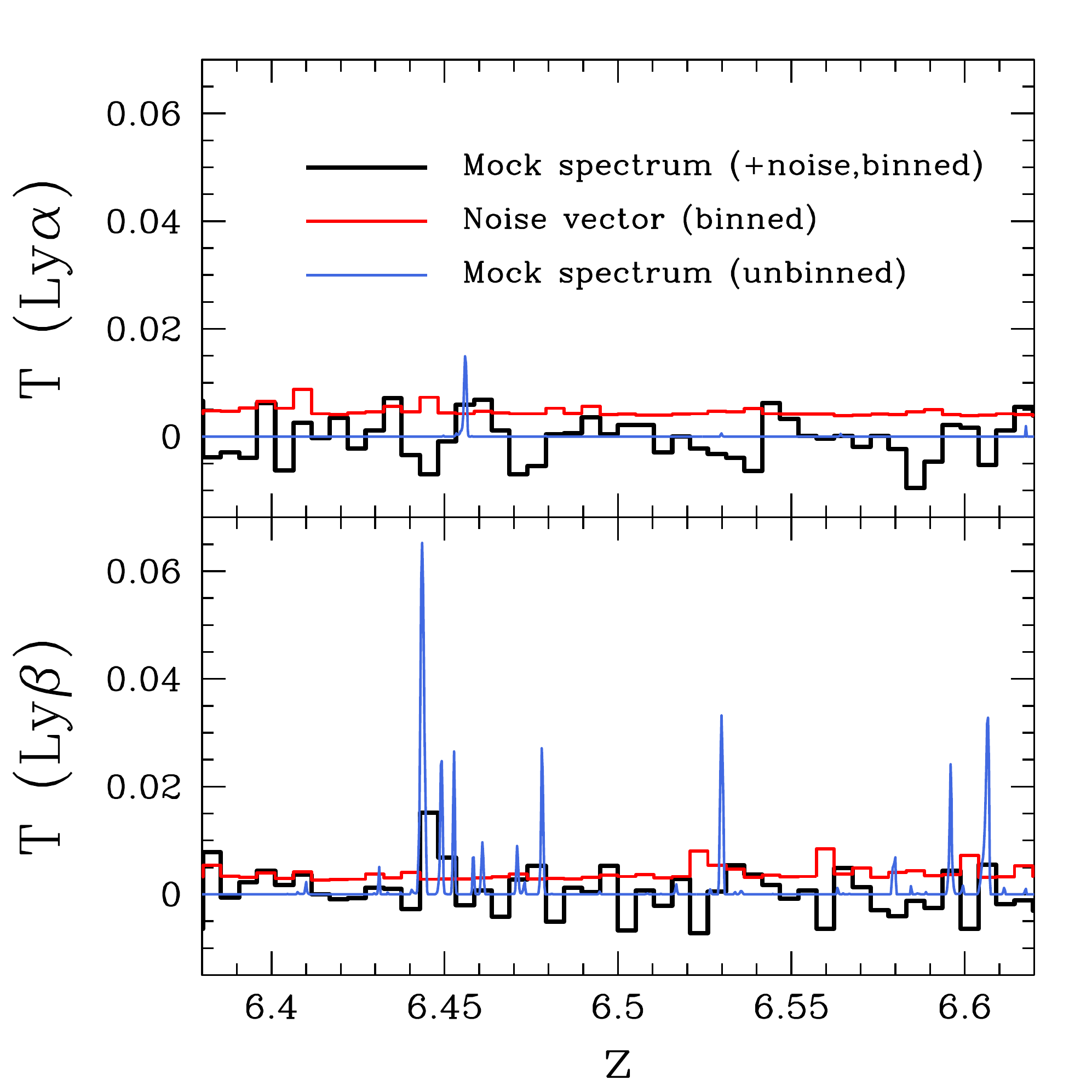}
\caption{Top panels: Mock \lya and \lyb forest spectra (2 Mpc binned: black, noiseless unbinned: blue) and $1\sigma$ noise (2 Mpc binned, red) for $\GHI=10^{-12.95}$ s$^{-1}$ at $z\sim6.5$. The mean transmissions of \lya and \lyb are consistent with zero, but there is a highly significant transmission spike in the \lyb forest which should definitively rule out $\GHI=0$.}
\label{fig:bestcase_data}
\end{figure}

\begin{figure}
\epsscale{1.1}
\plotone{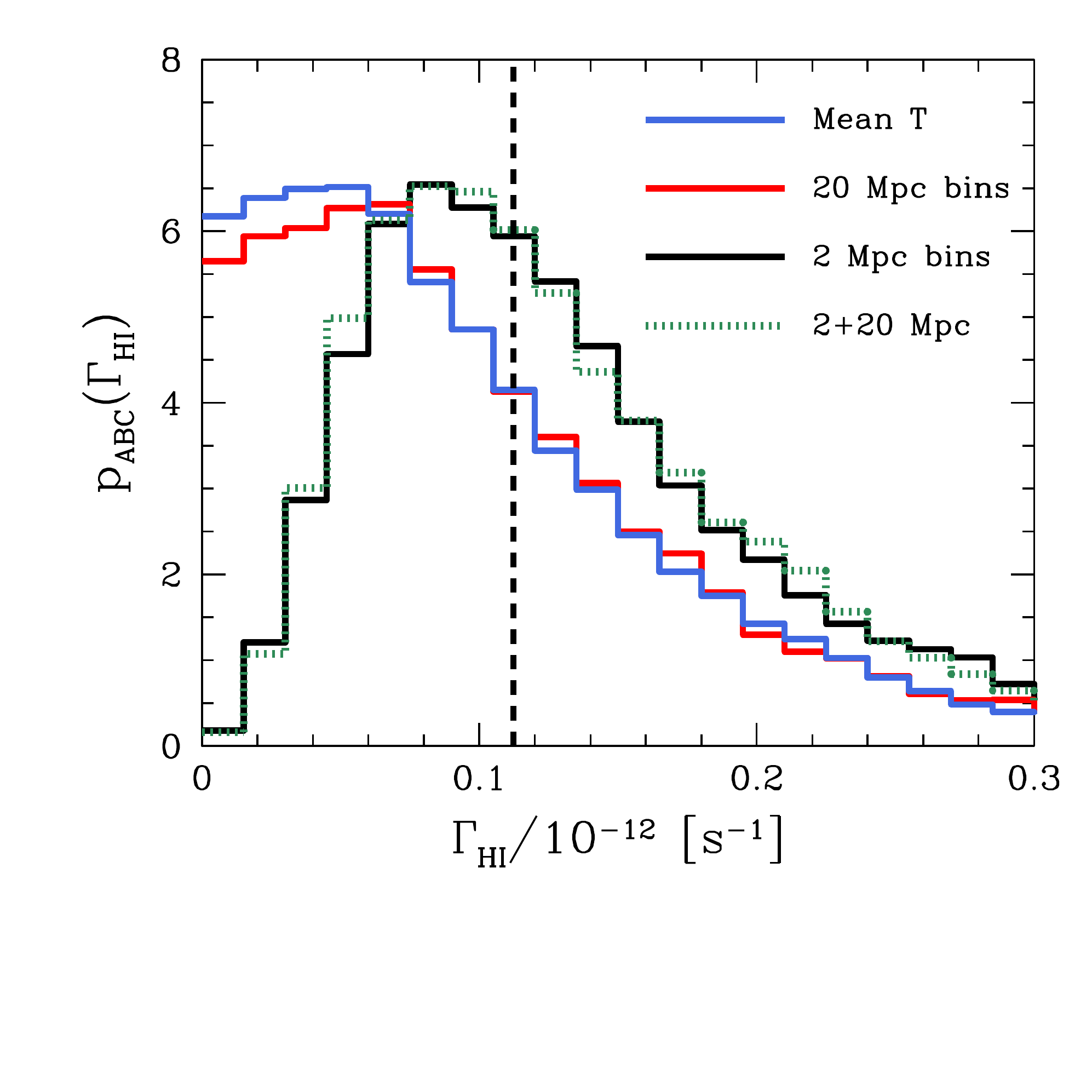}
\vskip -5em
\caption{Posterior PDFs of $\GHI$ for the \lya and \lyb forest spectra in Figure~\ref{fig:bestcase_data} using the mean transmission (blue) versus the rank-order distribution of 2 Mpc and 20 Mpc binned transmission (red and black, respectively) as the ABC summary statistic, compared to the true value (dashed line). While the mean transmission and 20 Mpc bins only provide an upper limit to $\GHI$, the S/N$\,\sim5.5$ 2 Mpc binned spike in \lyb is highly constraining. The green dotted curve shows the posterior PDF when the 2 and 20 Mpc rank-order distribution summary statistics are combined.}
\label{fig:bestcase_pdfs_sum}
\end{figure}

\begin{figure}
\epsscale{1.1}
\plotone{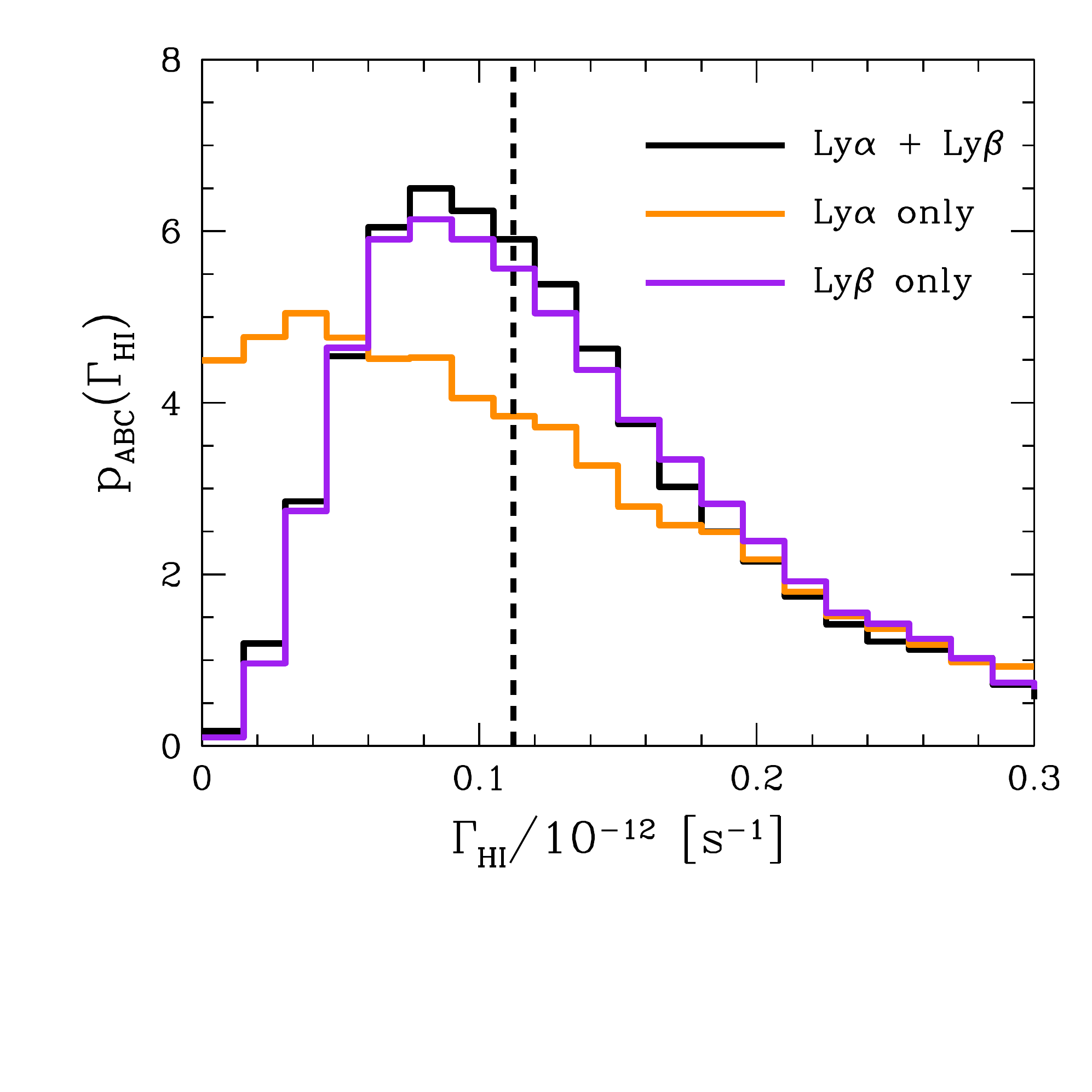}
\vskip -5em
\caption{Posterior PDFs of $\GHI$ for the \lya and \lyb forest spectra in Figure~\ref{fig:bestcase_data} using the rank-order distribution of 2 Mpc binned transmission in \lya alone (orange), \lyb alone (purple), and both (black) as the ABC summary statistic. The dashed line indicates the true value of $\GHI$.}
\label{fig:bestcase_pdfs_lyab}
\end{figure}

In Figure~\ref{fig:bestcase_data} we show a 2 Mpc-binned mock \lya and \lyb forest spectrum at $z\sim6.5$ with a fluctuating ionizing background and true $\GHI=10^{-12.95}$ s$^{-1}$. The mean transmissions in \lya and \lyb are consistent with zero, but there is a highly significant ($S/N>5$) transmission spike in the \lyb forest. In Figure~\ref{fig:bestcase_pdfs_sum}
we show posterior PDFs for $\GHI$ using several different summary statistics. The blue and red curves shows the posterior PDF using the mean transmission and the 20 Mpc-binned rank-order distribution as summary statistics, both of which fail to measure $\GHI$, i.e. the posterior PDF does not rule out $\GHI=0$. In contrast, using the 2 Mpc-binned rank-order distribution as the summary statistic (black curve) allows for an actual measurement of $\GHI$, demonstrating the enhanced ability of small-scale transmission features to constrain $\GHI$ relative to the large-scale mean. 

As an example of the flexibility of the ABC procedure, the dotted curve in Figure~\ref{fig:bestcase_pdfs_sum} shows the posterior PDF from adopting a summary statistic that \emph{combines} the 2 Mpc and 20 Mpc rank-order distributions by adding their distances in quadrature, i.e. $\rho_{\rm 2 Mpc+20 Mpc} = \sqrt{\rho_{\rm 2 Mpc}^2+\rho_{\rm 20 Mpc}^2}$. In Figure~\ref{fig:bestcase_pdfs_lyab}, we show the posterior PDFs using the 2 Mpc rank-order distribution summary statistic of the \lya forest alone (orange) and the \lyb forest alone (purple) compared to the joint constraints (black, same as Figure~\ref{fig:bestcase_pdfs_sum}). While the \lyb forest has the most constraining power due to the presence of the significant transmission spike, including the \emph{non-detection} of transmission in \lya modestly improves the posterior PDF density at the location of the true $\GHI$.

As an alternative to the rank-order distribution, we have also devised a one-dimensional summary statistic that is sensitive to
rare transmission spikes which we call the ``pseudo-likelihood" of the
spectrum. Through forward modeling many mock data sets,
we can compute the
transmission PDF in our simulations to arbitrarily high precision
(e.g. Fig.~\ref{fig:pdfs}) with any set of model parameters, and then compute
the ``pseudo-likelihood" of the spectrum under the
assumption that each bin of the \lya and \lyb
forests is
entirely independent: $p({\bf d}|\theta) = \prod_i
P(F_i|\theta)$. If there were no spatial correlations in (or between) the \lya and \lyb forests, this likelihood would be exact.
After finding the maximum pseudo-likelihood model
parameters $\theta_\mathrm{ML}$ of the observed data, we then use the
absolute difference of pseudo-likelihood values in log space as the
distance metric between the observed and mock data,
\begin{eqnarray}
&&\rho({\bf s}_\mathrm{obs},{\bf s})= \nonumber \\ && \left|\sum_i^{N_{\alpha+\beta}} \ln{P(F_{i,\mathrm{obs}}|\theta_\mathrm{ML})} - \sum_i^{N_{\alpha+\beta}} \ln{P(F_{i,\mathrm{mock}}|\theta_\mathrm{ML})}\right|,
\end{eqnarray}
where $i$ encompasses all of the bins in the \lya and \lyb forests and $N_{\alpha+\beta}=N_\alpha+N_\beta$. The pseudo-likelihood summary statistic described above is similar to ``indirect inference" ABC methodologies \citep[e.g.][]{Drovandi11,GP13}.
We find that this summary statistic is equally powerful as the rank-order distribution in terms of its ability to constrain $\GHI$ from sparse transmission spikes.

\begin{figure}
\epsscale{1.1}
\plotone{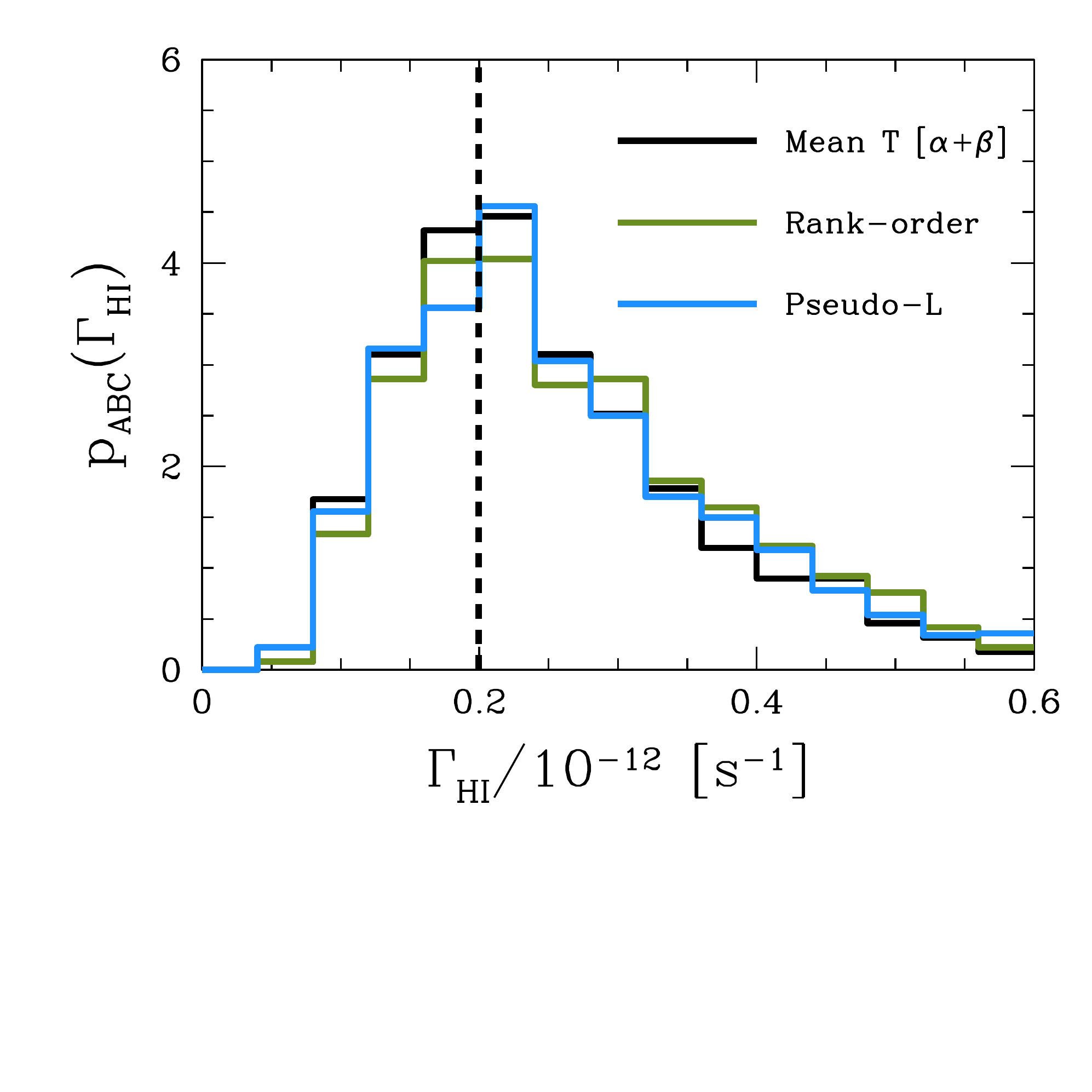}
\vskip -6em
\caption{Comparison of the ABC posterior distributions of $\GHI$ for the same mock spectrum as in Figure~\ref{fig:abcdemo} using three different summary statistics: mean transmission (black), rank-order distribution (green), and pseudo-likelihood (blue). The true $\GHI$ is shown as the vertical dashed line.}
\label{fig:sumdemo}
\end{figure}

Outside of examples like Figure~\ref{fig:bestcase_data}, the mean transmission, rank-order distribution, and pseudo-likelihood summary statistics are comparable in their ability to constrain $\GHI$. We demonstrate this in Figure~\ref{fig:sumdemo} where we show ABC posterior distributions for the three statistics computed for the same mock data as Figure~\ref{fig:abcdemo}, and we have chosen the tolerance $\epsilon$ such that we accept 5\% of the 25,000 mock datasets. The rank-order distribution and pseudo-likelihood statistics should be more sensitive to additional parameters that affect the shape of the transmission PDF such as the thermal state of the IGM \citep[e.g.][]{Lee15} and the strength of ionizing background fluctuations \citep[e.g.][]{MW03}. We restrict ourselves to inferring $\GHI$ alone in this work for simplicity in this initial presentation of our methodology, and we note that this choice limits the additional information gathered by the distribution of \lya and \lyb forest transmission. That is, in the models we present in this work there is a one-to-one relationship between the mean transmission and $\GHI$, and the scale-dependent distribution of transmission does not contain much additional information \emph{unless} the mean transmission is close to the noise limit of the spectrum. The methodology in principle allows for measurements of other parameters, such as the strength and scale of $\GHI$ fluctuations and the thermal state of the IGM, which we will pursue in future work.

\subsection{Validation of Statistical Methods: Inference Tests}\label{sec:inftest}

\begin{figure}
\epsscale{1.2} \plotone{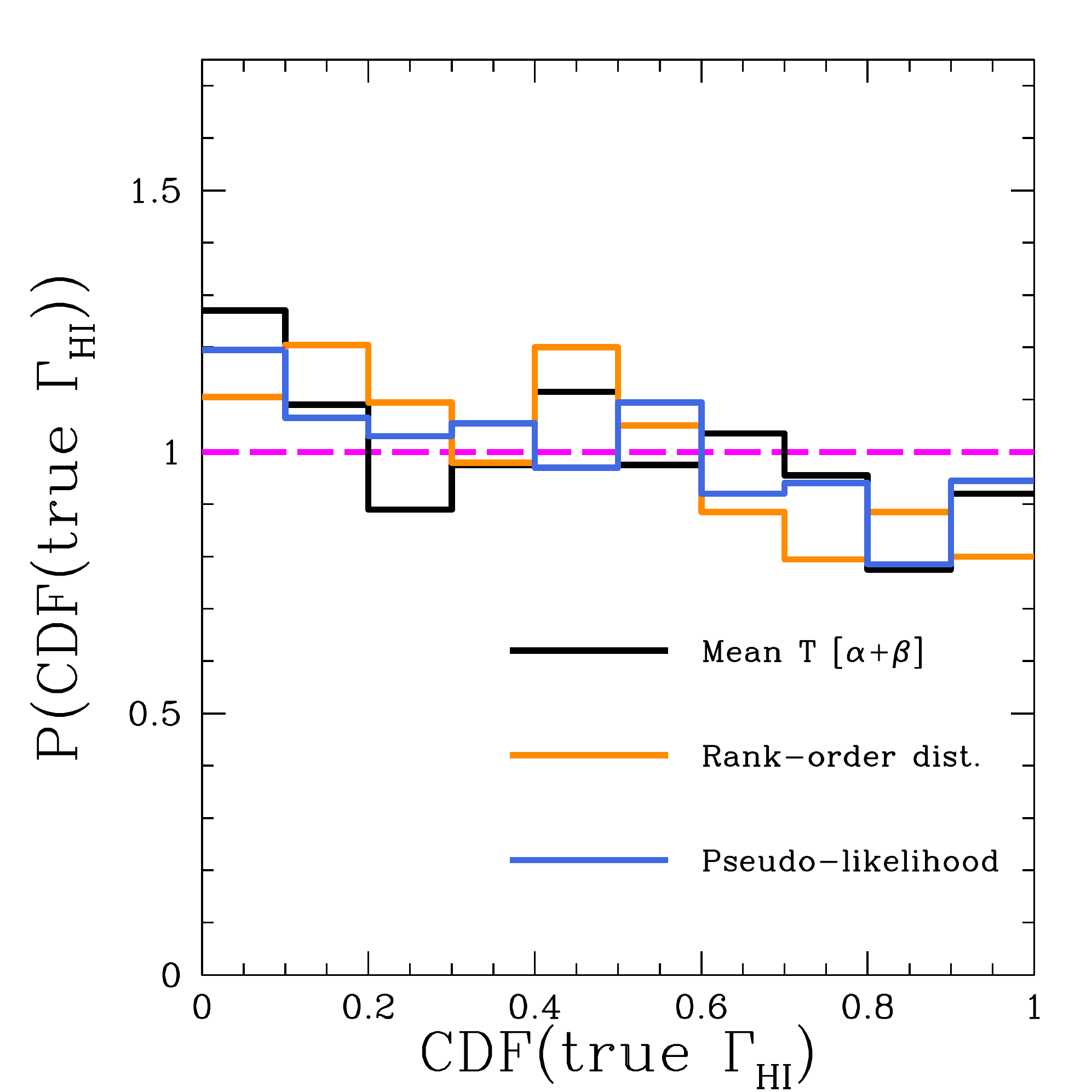}
\caption{Inference test of the ABC methodology using the different summary statistics described in \S~\ref{sec:abcapp}. A total of 2000 mock measurements of $\GHI$ were performed on 2000 different mock data sets (consisting of one noisy \lyab\ forest spectrum covering 100 Mpc) with $\GHI=10^{-12.7}$ s$^{-1}$. The solid curves show the distribution of the posterior CDF of each mock measurement evaluated at the true $\GHI$ (i.e. CDF(true $\GHI$) as defined in the text) using the mean transmission (black), rank-order PDF (orange), and pseudo-likelihood (blue) as summary statistics.} 
\label{fig:inftest}
\end{figure}

We rigorously test the capability of ABC to perform statistical inference of $\GHI$ as follows. We conduct
2000 mock $\GHI$ measurements similar to the examples in the previous section -- for each measurement,
we draw one mock dataset (100 Mpc of the \lya and \lyb forests at $z\sim6$ with $\GHI=10^{-12.7}$ s$^{-1}$, e.g. Figure~\ref{fig:mockspec})
to be the ``data", then draw 500 samples from the ABC approximate posterior distribution of $\GHI$.
For each of the 2000 mock measurements, we record the value of the cumulative distribution function (CDF), determined from the 500 posterior samples, at the true $\GHI$, CDF(true $\GHI$). The procedure then results in 2000 values of CDF(true $\GHI$). 

The simplest inference test would be to test what fraction of the time, out of the 2000
mock measurements,
$0.16 <$ CDF(true $\GHI) < 0.84$ (central 68\%), or $0.025 <$ CDF(true $\GHI) < 0.975$ (central 95\%), etc., but we opt for a somewhat stricter test covering the entire CDF. If the ABC posterior distribution can be treated as a true probability distribution, then the probability that the true $\GHI$ lies within \emph{any} interval of the CDF should be equal to the size of that probability interval, i.e. 
\begin{equation}
P(A < \mathrm{CDF}(\mathrm{true}\ \GHI) < B) = B-A.
\end{equation}
In other words, the distribution of CDF(true $\GHI$) should follow the uniform distribution over the interval [0,1].

We demonstrate this ``CDF(true $\GHI$)" test for the mean transmission, rank order distribution, and pseudo-likelihood summary statistics in Figure~\ref{fig:inftest} for CDF(true $\GHI$) bins of width $\Delta {\rm CDF}=0.1$. The dashed line shows the expected distribution of CDF(true $\GHI$) (i.e. a uniform distribution), while the solid curves show the distribution of CDF(true $\GHI$) of the 2000 mock measurements of \lya and \lyb forest spectra described above. In general, the ABC posterior distributions behave similar to the expectations for a true probability distribution, but there is a small tilt such that CDF(true $\GHI$) tends to be $<0.5$ more often than it should (i.e. P(CDF(true $\GHI$) $< 0.5 \sim 0.53$--$0.56$). This deviation from a uniform distribution is likely due to a combination of factors. First, the summary statistics may be insufficient, leading to bias in the posterior distribution. Second, the posterior PDFs from each test may not be fully converged due to the fact that $\epsilon>0$, i.e. the tilt demonstrates the inherent ``approximate" nature of the ABC posterior. 
Finally, we drew only 500 posterior samples for each test, and this may not be enough to adequately populate the tails of the posterior PDF. Nevertheless, this rigorous test shows that the probabilities from the ABC posterior distribution can be reasonably interpreted as ``true" probabilities, at least in the context of our theoretical model for the IGM.

\section{Proof of concept $\GHI$ measurement}\label{sec:poc}

\begin{figure}
\resizebox{8.8cm}{!}{\includegraphics[trim={5.5em 27em 0 0},clip]{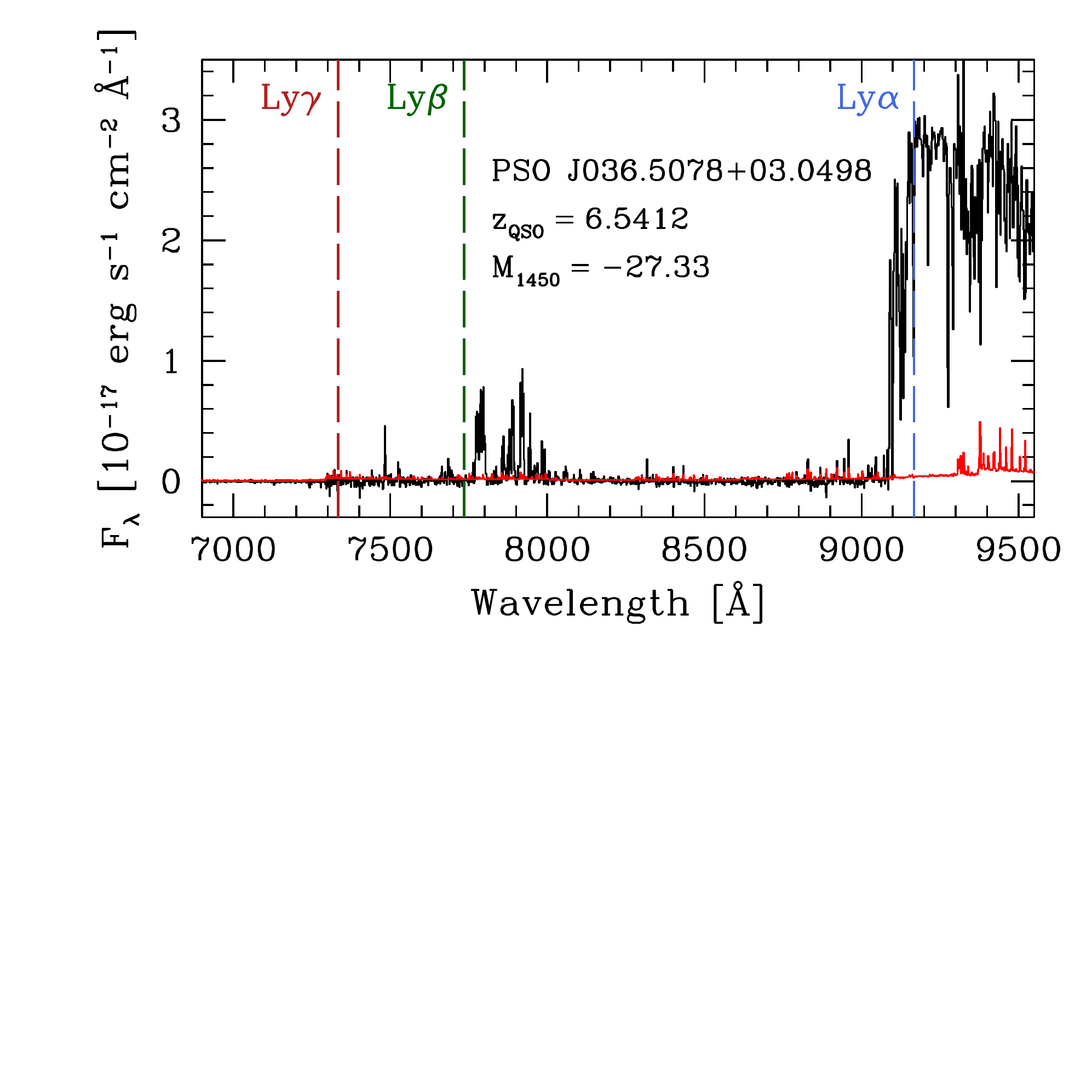}}
\caption{Keck/ESI spectrum of PSO J036.5078+03.0498 ($z_\mathrm{QSO}=6.5412$). The spectrum (black) and noise vector (red) have been binned by six pixels ($\approx1$ resolution element). The wavelengths of $\lyam$, $\lybm$, and Ly$\gamma$ at $z_\mathrm{QSO}$ are denoted by vertical lines.}
\label{fig:j0226spec}
\end{figure}

\begin{figure*}
\resizebox{18cm}{!}{\includegraphics[trim={4em 30em 0 0},clip]{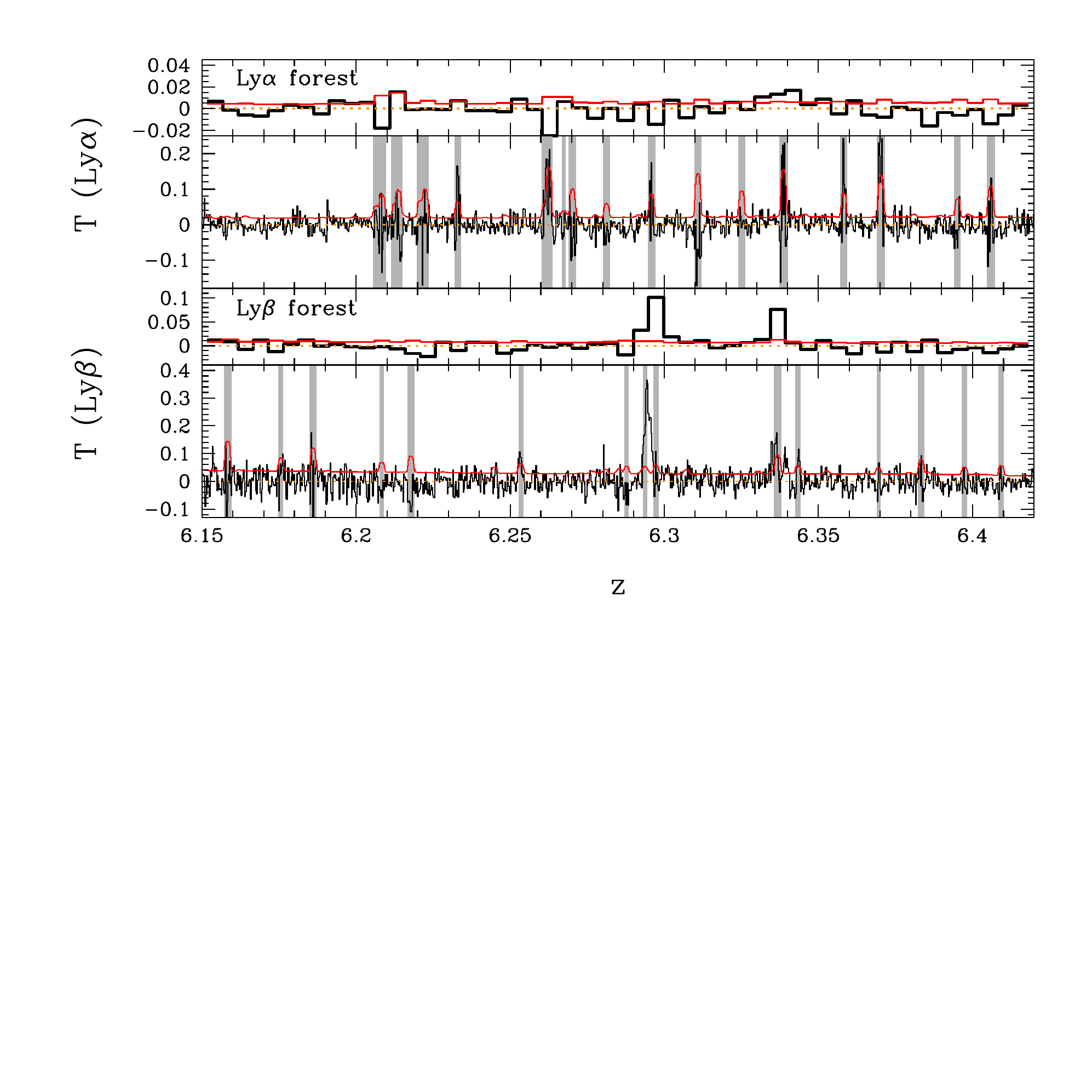}}
\caption{Overlapping \lya and \lyb forest in the highest redshift bin studied in our proof-of-concept test on P036+03. The top two panels show the \lya forest, while the bottom two panels show the \lyb forest. The lower panel in each pair shows the un-binned spectrum (black) and noise vector (red), with grey shading indicating regions with strong sky lines that we have masked in our current analysis with an automated threshold above the intra-line noise level. The upper panel in each pair shows the 2 Mpc binned spectrum (black) and noise vector (red) only including pixels outside of the masking in the lower panels.}
\label{fig:j0226zoom}
\end{figure*}

As a proof of concept, we have applied our statistical methodology to a Keck/ESI spectrum of the recently discovered quasar PSO J036.5078+03.0498 \citep[$z=6.5412$,][henceforth P036+03]{Venemans15a,Banados15}, shown in Figure~\ref{fig:j0226spec}. The quasar was observed on January 11th and 12th 2016 for a combined exposure time of 11,700s with a slit width of 1", resulting in a signal-to-noise ratio of $\sim 20$--$45$ per pixel between sky lines in the extrapolated continuum level in the \lya and \lyb forests and a spectral resolution of $R\sim4000$. In the following analysis, and for the purposes of generating mock spectra, we use the noise vector from this observation instead of the one described in \S~\ref{sec:sims}. Further details of the data reduction and continuum fitting of this spectrum will be discussed in detail by Eilers et al. (in prep.).
The usable \lya forest of this quasar spans $z\sim5.36$--$6.42$, bounded by the onset of \lyb absorption at $(1+z)=(1+z_\mathrm{QSO})(\lambda_\lybm/\lambda_\lyam)$ and by the $\Delta z\sim0.1$ extent of the quasar proximity zone (as defined by the declining visual excess in transmission spikes, cf. \citealt{Fan06}). The usable \lyb forest spans $z\sim6.15$--$6.42$, similarly cut off by the onset of \lyg absorption at $(1+z)=(1+z_\mathrm{QSO})(\lambda_\lygm/\lambda_\lygm)=6.15$ and by the quasar proximity zone at $z=6.42$.

In Figure~\ref{fig:j0226zoom}, we show a zoom-in on the overlap between the \lya and \lyb forests at $z\sim6.3$. Due to the presence of correlated noise in sky lines that we currently ignore in our modeling procedure, we have automatically masked regions which have noise greater than twice the smoothly varying inter-line noise, shown as the grey vertical bands. The \lya forest is nearly consistent with zero transmission, but when binned to 2 Mpc ($\sim20$ ESI pixels) there is a region with S/N$\,\sim3$ at $z\approx6.34$.
The \lyb forest shows considerably more transmission, mostly notably a strong \lyb spike at $z\approx6.295$ which is partially masked due to its overlap with sky lines. A weaker \lyb feature also appears at $z\approx6.335$ which is similarly contaminated by sky lines, and may be due to the same physical structure (i.e. void in the IGM) as the \lya feature at the same redshift.

\begin{figure}
\epsscale{1.2}
\plotone{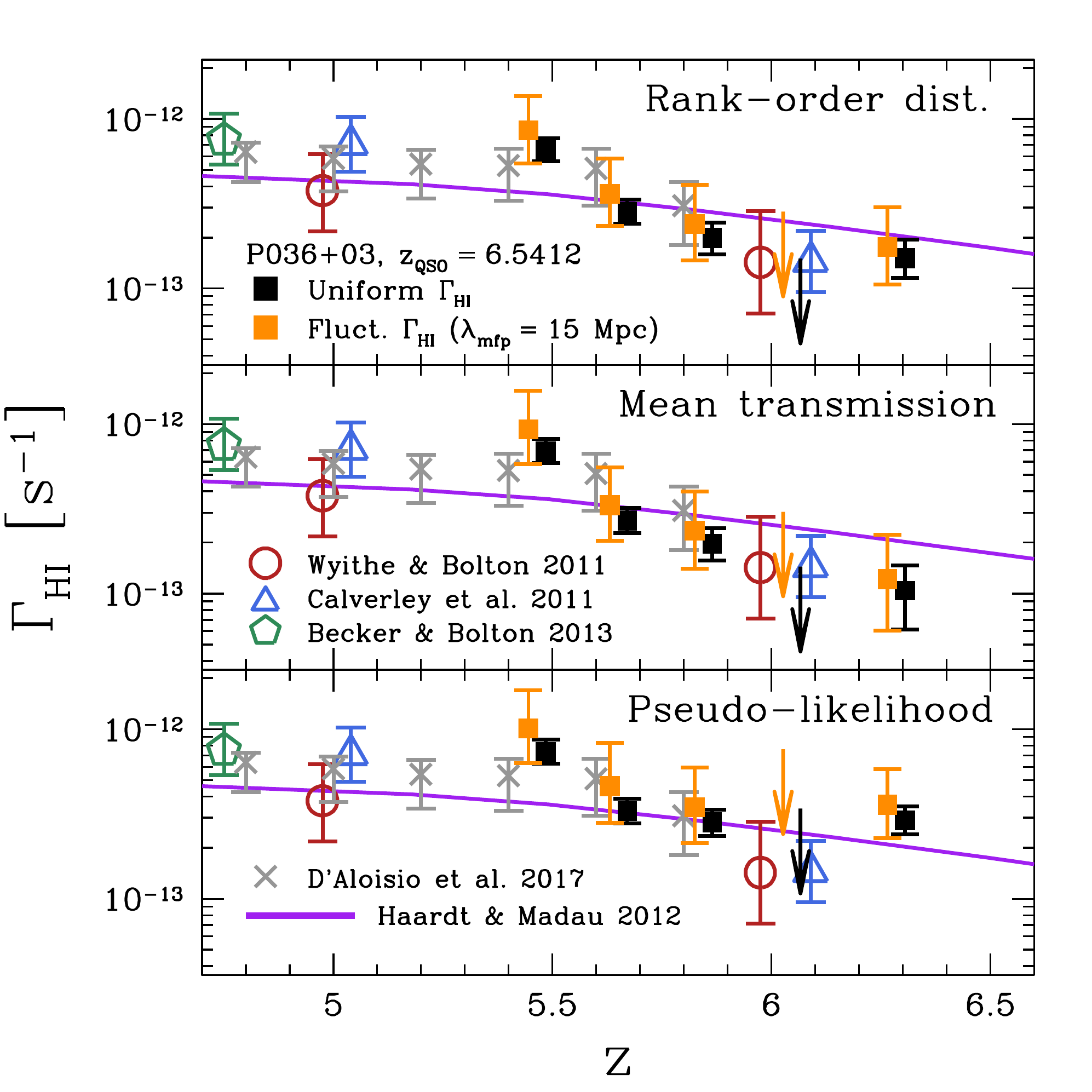}
\caption{Proof of concept measurements of $\GHI$ from the transmission spectrum of P036+03 shown in Figure~\ref{fig:j0226spec} divided into the bins listed in Table 1, compared to previous measurements and the commonly assumed Haardt \& Madau (2012) model. The black and orange points show the median $\GHI$ of the posterior PDFs in each redshift bin assuming uniform and fluctuating ionizing background ($\lambda_\mathrm{mfp}=15$ Mpc) models, respectively. Error bars denote the 16th-84th percentile range of the posterior PDFs. Where a measurement is consistent with zero, we show a downward arrow at the 95th percentile of the posterior PDF. The top, middle, and bottom panels apply the rank-order distribution, mean transmission, and pseudo-likelihood summary statistics, respectively.}
\label{fig:j0226poc}
\end{figure}

\begin{figure}
\resizebox{8.8cm}{!}{\includegraphics[trim={6em 28em 0 -1em},clip]{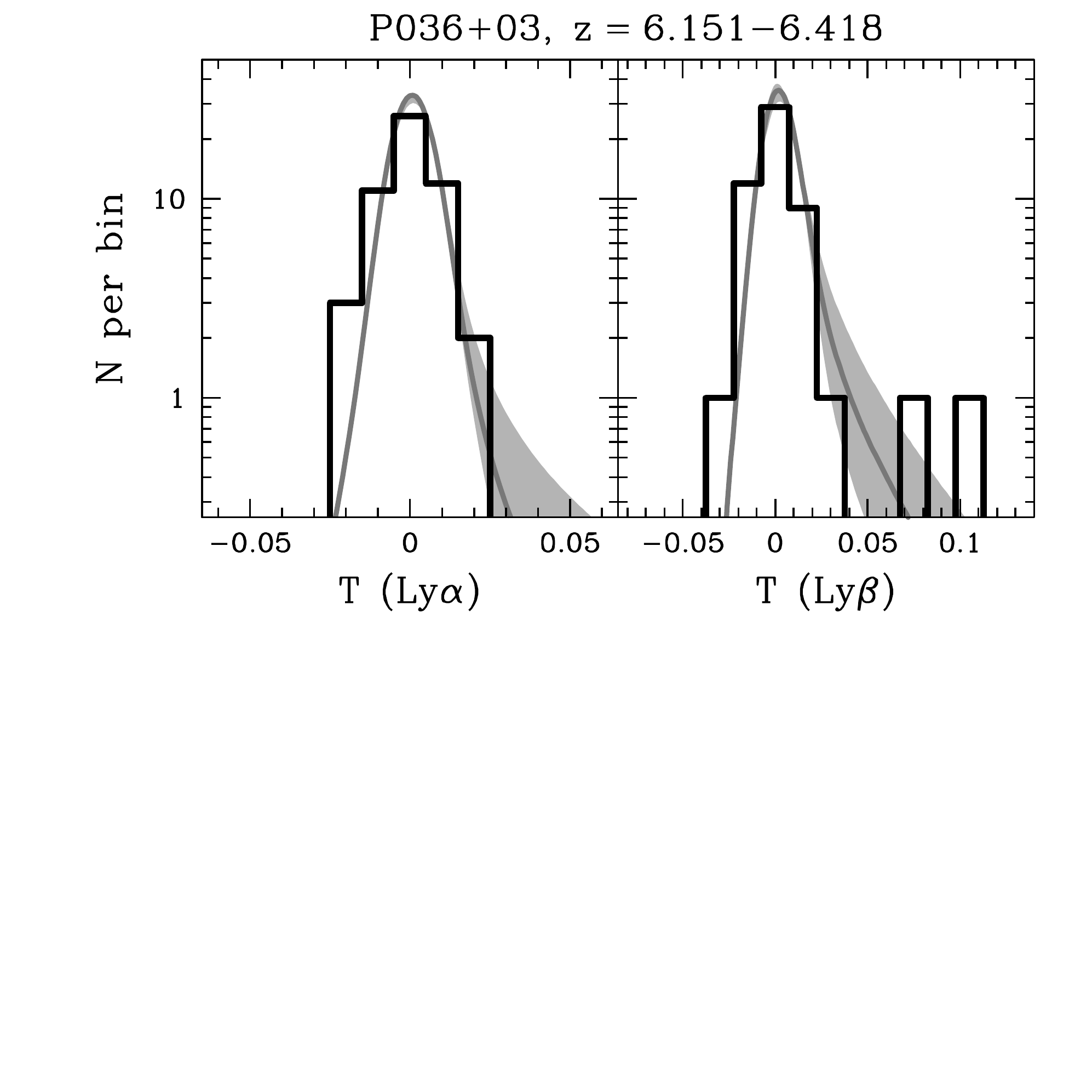}}
\caption{Transmission PDFs of the 2 Mpc-binned \lya (left) and \lyb (right) forest of P036+03 at $z=6.151$--$6.420$. The black histograms show the measured distribution of transmission for P036+03, while the grey curve and shaded regions show (fluctuating background) model PDFs corresponding to the median and 68\% credible $\GHI$ values, respectively, as measured by the rank-order distribution summary statistic assuming a fluctuating ionizing background (Table 1).}
\label{fig:j0226pdf}
\end{figure}

\begin{figure}
\resizebox{8.8cm}{!}{\includegraphics[trim={0 0 0 0},clip]{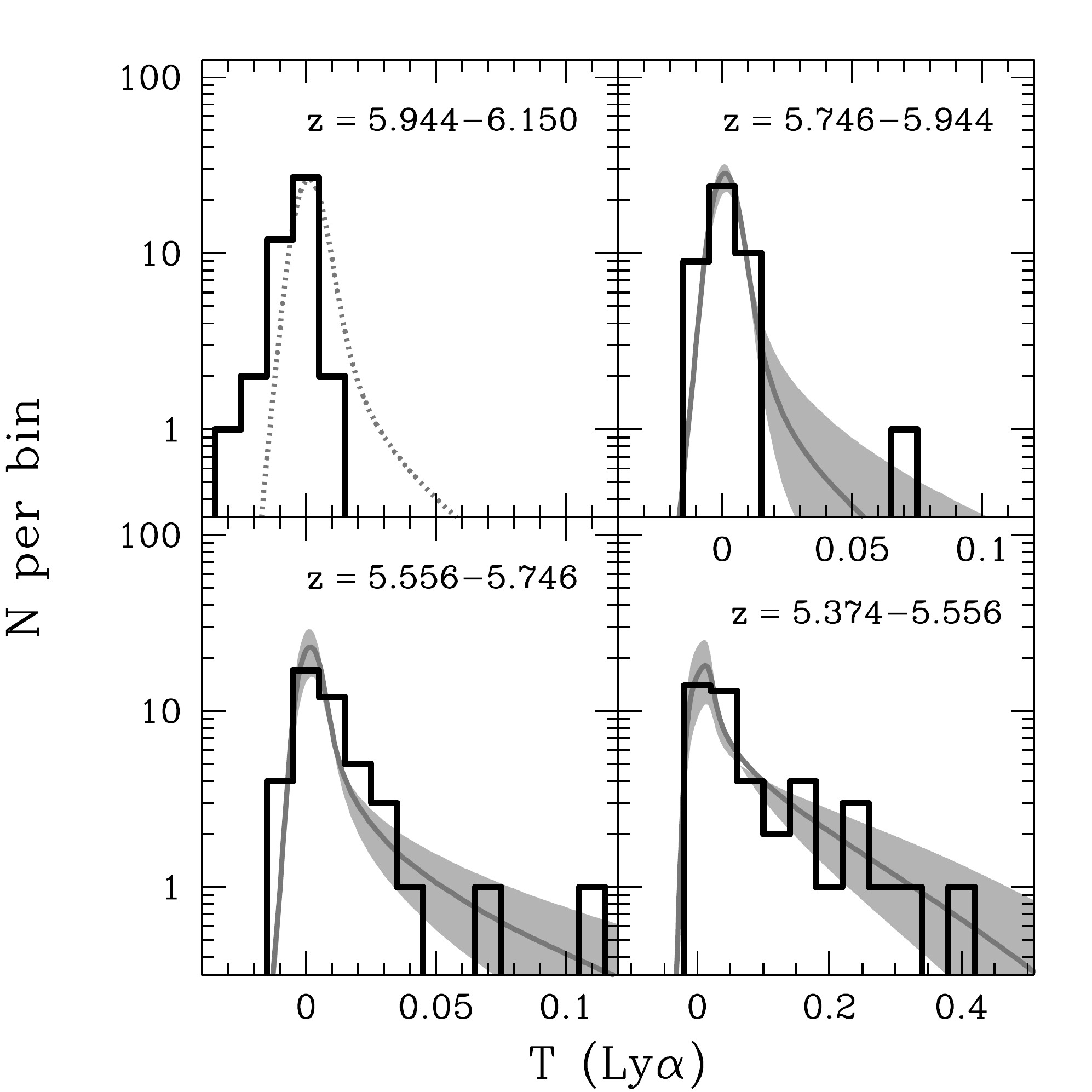}}
\caption{Transmission PDFs of the 2 Mpc-binned \lya forest of P036+03 in redshift bins below the \lyb forest overlap with corresponding (fluctuating background) model PDFs similar to Figure~\ref{fig:j0226pdf}. The dotted curve in the upper left panel reflects the model PDF corresponding to the 95\% credible upper limit for $\GHI$ given the lack of significantly detected transmission. Note the extended horizontal axis scale on the lower right panel.}
\label{fig:j0226gallery}
\end{figure}

We have divided the \lya forest into the five redshift bins listed in Table~1. The highest redshift bin covers 108 Mpc corresponding to the entire \lyb forest range,
while the remaining four bins split the forest into 88 Mpc
regions covering the rest of the \lya forest. In each redshift bin, we perform separate ABC analyses using the three summary statistics described in \S~\ref{sec:abcapp}: the mean transmission, the rank-order distribution of 2 Mpc pixels, and the pseudo-likelihood of 2 Mpc pixels.
The same pixels in the mock data sets were masked as in the observed spectra so as to keep the analysis of the mocks similar to what was applied to the real data.
We assume a broad uniform prior on $\GHI$ from 0--$3\times10^{-12}$ s$^{-1}$ and draw 2500 samples from the (approximate) posterior PDF of $\GHI$.

\begin{table}
\begin{center}
\caption{Proof-of-concept $\GHI$ measurements for P036+03}
\label{tab:j0226}
\begin{tabular}{c c c c c}
\hline \noalign {\smallskip}
$z$ & $dR^a$ & Ly-series & $\log{\GHI^{\mathrm{unif}}}$ (s$^{-1}$) & $\log{\GHI^{\mathrm{fluct}}}$ (s$^{-1}$) \\
\hline \noalign {\smallskip}
6.151--6.418 & 108 & Ly$\alpha$+Ly$\beta$ & $-12.82^{+0.11}_{-0.11}$ & $-12.76^{+0.23}_{-0.22}$ \\
5.944--6.150 & 88 & \lya\ & $(-12.83)^b$ & $(-12.55)^b$ \\
5.746--5.944 & 88 & \lya\ & $-12.70^{+0.09}_{-0.10}$ & $-12.62^{+0.23}_{-0.20}$ \\
5.556--5.746 & 88 & \lya\ & $-12.55^{+0.07}_{-0.07}$ & $-12.44^{+0.21}_{-0.19}$ \\
5.374--5.556 & 88 & \lya\ & $-12.18^{+0.06}_{-0.07}$ & $-12.07^{+0.22}_{-0.19}$ \\
\hline \noalign {\smallskip}
\end{tabular}
\end{center}

The reported values of $\GHI$ are the medians of the posterior PDFs, while the error bars represent the central $68\%$ credible interval. The posterior PDFs were computed using ABC with the rank-order distribution summary statistic. Note that these uncertainties \emph{do not} include systematic uncertainties due to the assumed thermal state of the intergalactic medium or the continuum fit of the quasar spectrum.

$^a$ Size of redshift range in Mpc

$^b$ 95\% credible upper limits.
\end{table}

Figure~\ref{fig:j0226poc} shows the resulting $\GHI$ measurements and uncertainties compared to previous measurements. We show two limiting cases of the ionizing radiation field -- a uniform background (black points), and a highly fluctuating background as suggested by DF16 at $z\ga5.6$ (orange points). The fluctuating background models assume the $\GHI/\langle\GHI\rangle$ field computed as described in \S~\ref{sec:sims}
with $\lambda_\mathrm{mfp}=15$ Mpc (e.g. bottom panel of Figure~\ref{fig:mockspec}, drawn from the simulation shown in Figure~\ref{fig:uvb}), and the constraints in Figure~\ref{fig:j0226poc} apply to $\langle\GHI\rangle$. In general the measurements are consistent with previous work at $z\sim5$--$6$ \citep[large open points;][]{WB11,Calverley11,BB13,D'Aloisio17}.
These works measured $\GHI$ two different ways, either by calibrating hydrodynamical simulations to measurements of the mean transmission \citep{WB11,BB13,D'Aloisio17} or by analyzing the transmission profile of quasar proximity zones \citep{Calverley11}. For those measurements based on the mean transmission we have adjusted their $\GHI$ values slightly to account for their different assumed cosmologies following the scaling factors in \citet{BH07a}. Each of these measurements represents samples of $\geq7$ quasars, while our comparably precise measurements are from a single quasar, and we more explicitly model large-scale cosmic variance of the density field and a strongly fluctuating ionizing background. However, our stated uncertainties do not include significant systematic uncertainties considered by previous authors, which we discuss further below.

In Figure~\ref{fig:j0226pdf} we show the measured PDFs of 2 Mpc transmission ($dT=0.01$-binned) in the \lya and \lyb forests at $z=6.151$--$6.420$  compared to fluctuating background model PDFs corresponding to the median and 68\% credible
posterior values of $\GHI$ measured by the rank-order distribution (Table 1). The constraints demonstrate the capability of the ABC procedure to constrain $\GHI$ from a sparsely sampled tail of significant transmission measurements. In Figure~\ref{fig:j0226gallery} we show similar plots for the remaining lower redshift bins which are constrained by the \lya forest alone. The upper left panel shows the bin of $z=5.943$--$6.151$ over which there are no significant transmission spikes detected, with the model PDF corresponding to the 95\% credible upper limit on $\GHI$ shown by the dotted curve. In this redshift range, corresponding to observed wavelengths $8440$--$8693$\AA, the excess of pixels with relatively large negative transmission values compared to the model PDF suggests that there are systematics in the data reduction that are not present in our simple Gaussian noise model. The other redshift bins show good agreement between the shape of the model PDFs and the data.

The constraints from different summary statistics are nearly identical, although the pseudo-likelihood statistic leads to somewhat elevated $\GHI$ at all redshifts, most prominently in the highest redshift bin. This is likely due to our simplistic modeling of the noise in the spectrum as uncorrelated and Gaussian-distributed. In future work, we will use more realistic noise realizations extracted from real data. Both the pseudo-likelihood and mean transmission statistics are adversely affected by regions with an excess of negative transmission -- which exist in the data but not in our noise model -- leading to positive and negative biases in the $\GHI$ measurements, respectively. 

In detail, the pseudo-likelihood statistic cannot distinguish between
negative and positive outliers because the negative side of the
transmission PDF is only weakly dependent on
$\GHI$
(e.g. Figure~\ref{fig:pdfs}), so excess negative transmission
(wrongly) influences the posterior PDF the same way as a positive
transmission spike, e.g. towards higher $\GHI$.
Conversely, excess
negative transmission decreases the mean transmission more than
otherwise expected by our assumed noise variations, so the inferred
$\GHI$ can be biased low. Despite the presence of small-scale
transmission in the \lya and \lyb forests in the highest redshift bin
(Figure~\ref{fig:j0226zoom}), the mean transmission in \lya is consistent
with zero, and the mean transmission in \lyb is only barely detected
with ${\rm S\slash N}\approx2$.

Note that, for simplicity, we have not included uncertainties in the
thermal state of the IGM nor uncertainties in the quasar continuum. In
the case of a uniform background, the former uncertainty dominates the
error budget of $\GHI$ \citep{BB13}. Preliminary tests marginalizing
over a wide range in potential IGM thermal states and $\pm20\%$
continuum uncertainty\footnote{Due to the weak transmission at
  $z\ga5.5$, uncertainties in the quasar continuum have a relatively
  small effect on the inferred ionizing background (see \citealt{Eilers17}). In the simplest
  case of inference via the mean transmission, $\GHI \propto
  1/\teff^{\sim2.0}$ \citep{BB13}. Propagating uncertainties through
  $\teff = -\ln{\langle F \rangle}$, a continuum error of 20\% implies
  $\sigma_{\teff}\sim0.2$, which then corresponds to
  $\sigma_{\GHI}/\GHI\sim$ 10\% for $\teff\sim4$.}
suggest that the
true error budget should be inflated by up to $\sim0.2$ dex, but we
leave a detailed multi-parameter analysis to future work. We have also
ignored uncertainties in the mean transmission at $z\sim5$ which will
contribute uncertainty to the \lyb measurement -- however, the
uncertainty in the mean transmission is likely to be smaller than the
intrinsic scatter between sightlines due to the IGM density field,
which we explicitly model.
Thus the uncertainties shown in Figure~\ref{fig:j0226poc} and listed in Table~1 should be understood as lower limits based on statistical uncertainty alone.
Given our ability to measure $\GHI$ from individual sightlines, in principle we can determine how realistic our uncertainties are by comparing our predicted dispersion (from the posterior PDFs) to the actual dispersion in measurements on multiple sightlines. In future work we will directly compare our estimated uncertainties in the mean $\GHI$ to the dispersion in actual measurements from a large ensemble of quasars.

\begin{figure}
\epsscale{1.2}
\plotone{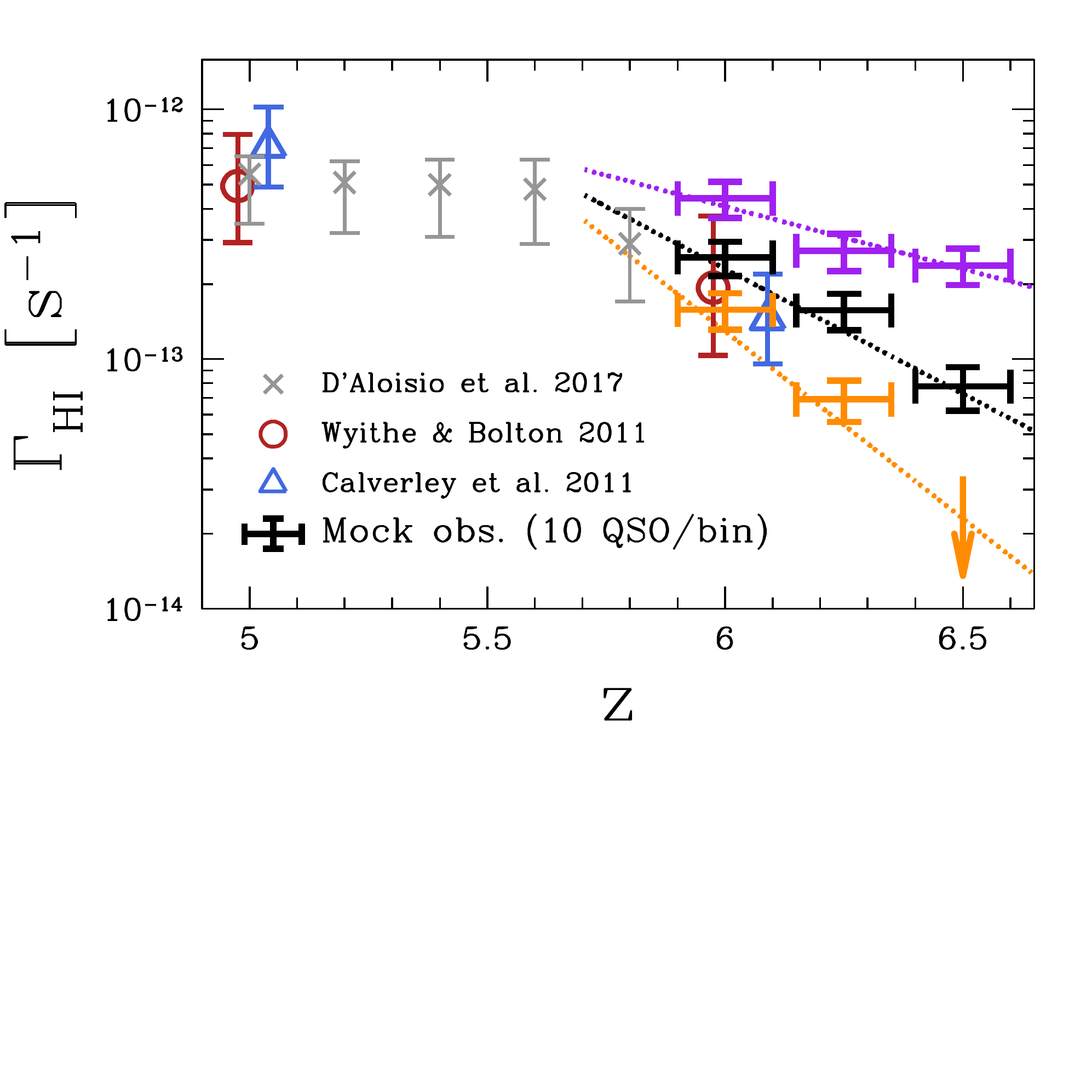}
\vskip -9em
\caption{Purple, black, and orange error bars demonstrate the median $\GHI$ of the posterior PDF from measurements performed on mock data sets (10 quasars per bin, using the overlapping \lya and \lyb forest region) following the evolving $\GHI$ models shown by the corresponding dotted curves. The downward arrow in the highest redshift bin of the lowest $\GHI$ model represents a $2\sigma$ upper limit from the lack of significant transmission spikes in the mock data.}
\vskip -0.1em
\label{fig:mocksurvey}
\end{figure}

\section{Discussion and Conclusion}\label{sec:end}

The $\sim$Mpc-scale distribution of transmission in the \lya and \lyb forests at $z\ga6$ contains considerable statistical constraining power for the ionization state of the post-reionization IGM, but extracting this information from observed spectra is challenging. We have shown that an application of ABC
should allow for precise measurements of the ionizing background at high redshift, even from $\sim 1$ significant transmission spike in a broad spectral region.

We have demonstrated above that our method, applied to a single high-quality quasar spectrum, allows for a measurement of $\GHI$ to a precision of roughly a factor of two at $z>6$ in the presence of a highly fluctuating ionizing background. This uncertainty. is dominated by cosmic variance, primarily due to the fluctuating ionizing background with a smaller contribution from density fluctuations
In a fixed redshift interval, the constraining power of the \lya and \lyb forests increases roughly as $\sqrt{N_\mathrm{QSO}}$, so larger samples will not only be able to test the consistency of our uncertainty estimates but also allow for unprecedented precision in $\GHI$ measurements.  The population of known $z\ga6$ quasars has greatly increased in the last few years from all-sky surveys such as SDSS \citep[e.g.][]{Jiang15} and Pan-STARRS \citep[][Mazzucchelli et al., in prep.]{Venemans15a,Banados16}, and other ongoing searches for $z\ga6$ quasars \citep[e.g.][]{Reed15,Reed17,Venemans15b} will likely increase the potential spectroscopic sample size even further. 

In Figure~\ref{fig:mocksurvey}, we show mock measurements of $\GHI$ for a hypothetical survey of three samples of ten quasars each with $z_\mathrm{Q}=6.25,6.50,6.75$ to measure $\GHI$ at $z_\mathrm{IGM}=6.00,6.25,6.50$, respectively,
using the rank-order distribution summary statistic on 100 Mpc of the overlapping \lya and \lyb forest regions.
The three colors/dotted curves correspond to different toy evolution models that roughly bracket published measurements of the \lya forest effective optical depth at $z\ga 6$ (Figure 1) and we assume a strongly fluctuating ionizing background (\mfp\ $=15$ Mpc). While we do not include them in Figure~\ref{fig:mocksurvey}, at redshifts below where the \lyb forest is cut off by the onset of \lyg absorption the constraints from the \lya forest alone (down to $z\sim5.25$ for quasars at $z_\mathrm{QSO}=6$)
would be of comparable precision.

Currently, there are $\sim20$ $z\ga6$ QSOs with moderate-resolution (e.g. Keck/ESI, VLT/X-Shooter) spectra of comparable quality (or better) to our spectrum of P036+03 in public  
data archives.
In future work, we will
perform a joint statistical analysis of all publicly available spectra
of $z\ga6$ quasars to measure $\GHI$ from $z\sim5.5$ to $z\ga6.5$ and constrain IGM thermal parameters.
The
lower panels of Figure~\ref{fig:pdfs} show that background
fluctuations can imprint substantial signals in the distribution of
transmission in the \lya and \lyb forests on small and large
scales. With a large number of independent sightlines through the IGM,
it may be possible to constrain not only the mean strength of the
ionizing background, but also the strength and scale of ionizing
background fluctuations through a multi-scale approach.

\section*{Acknowledgements}

We thank M. Fouesneau, D. Hogg, and D. Foreman-Mackey for helpful conversations regarding statistical methods and ABC. Calculations presented in this paper used resources of the National Energy Research Scientific Computing Center (NERSC), which is supported by the Office of Science of the U.S.~Department of Energy under Contract No.~DE-AC02-05CH11231.

The authors wish to recognize and acknowledge the very significant cultural role and reverence that the summit of Mauna Kea has always had within the indigenous Hawaiian community.  We are most fortunate to have the opportunity to conduct observations from this mountain.

\bibliographystyle{apj}
 \newcommand{\noop}[1]{}

\end{document}